\documentclass[onecolumn,floatfix,amsmath,amssymb,prb,superscriptaddress]{revtex4}
\usepackage{graphics}
\usepackage{graphicx}
\usepackage{bm}         
\usepackage{amsmath}
\date{\today}
\begin{document} 
\newcommand{\beq}{\begin{equation}}
\newcommand{\eeq}{  \end{equation}}
\newcommand{\ba}{\begin{eqnarray}}
\newcommand{\ea}{  \end{eqnarray}}
\newcommand{\ve}{\varepsilon}
\newcommand{\bcite}[1]{[{\onlinecite{#1}}]}

\title{Current-induced forces in mesoscopic systems: a scattering matrix
approach}

\author{Niels Bode} 
\affiliation{\mbox{Dahlem Center for Complex Quantum Systems and Fachbereich
Physik, Freie Universit\"at Berlin, 14195 Berlin, Germany}}

\author{Silvia Viola Kusminskiy} 
\affiliation{\mbox{Dahlem Center for Complex Quantum Systems and Fachbereich
Physik, Freie Universit\"at Berlin, 14195 Berlin, Germany}}

\author{Reinhold Egger} 
\affiliation{Institut f\"ur Theoretische Physik, Heinrich-Heine-Universit\"at,
D-40225 D\"usseldorf, Germany}

\author{Felix von Oppen}
\affiliation{\mbox{Dahlem Center for Complex Quantum Systems and Fachbereich
Physik, Freie Universit\"at Berlin, 14195 Berlin, Germany}}

\begin{abstract}
Nanoelectromechanical systems are characterized by an intimate connection between electronic and mechanical degrees of freedom. Due to the nanoscopic scale, current flowing through the system noticeably impacts the vibrational dynamics of the device, complementing the effect of the vibrational modes on the electronic dynamics. We employ the scattering matrix approach to quantum transport to develop a unified theory of nanoelectromechanical systems out of equilibrium. For a slow mechanical mode, the current can be obtained from the Landauer-B\"uttiker formula in the strictly adiabatic limit. The leading correction to the adiabatic limit reduces to Brouwer's formula for the current of a quantum pump in the absence of the bias voltage. The principal result of the present paper are scattering matrix expressions for the current-induced forces acting on the mechanical degrees of freedom. These forces control the Langevin dynamics of the mechanical modes. Specifically, we derive expressions for the (typically nonconservative) mean force, for the (possibly negative) damping force, an effective "Lorentz" force which exists even for time reversal invariant systems, and the fluctuating Langevin force originating from Nyquist and shot noise of the current flow. We apply our general formalism to several simple models which illustrate the peculiar nature of the current-induced forces. Specifically, we find that in out of equilibrium situations the current induced forces can destabilize the mechanical vibrations and cause limit-cycle dynamics. 
\end{abstract}

\maketitle
\section{Introduction}
\label{sec0}
Scattering theory has proved a highly successful method for treating coherent
transport in mesoscopic systems \bcite{NazarovBook}. Part of its appeal is
rooted in its conceptual
simplicity: transport through a mesoscopic object can be described in terms of
transmission and reflection of electronic waves which are scattered by a
potential. This approach was introduced by Landauer
 \bcite{LandauerIBM57,LandauerPhilMag70} and generalized by
B{\"u}ttiker {\it et al.} \bcite{ButtikerPRB85} and leads to their well-known formula for the
conductance of multi-terminal mesoscopic conductors. For time-dependent phenomena, scattering matrix
expressions have been obtained for quantum pumping
 \bcite{BrouwerPRBR98,AvronPRL01}, a process by which a direct
current is generated through temporal variations of relevant
parameters of the
system, such as a gate voltage or a magnetic field. The case
of pumping in an out-of-equilibrium, biased system has remained largely unexplored so
far \bcite{Entin-WohlmanPRB02,MoskaletsPRB05}. 

The purpose of the present paper is to further develop the scattering matrix approach into a simple, unifying formalism to treat nanoelectromechanical
systems (NEMS). The coupling between mechanical and electronic degrees of
freedom is the
defining characteristic of NEMS~\bcite{CraigheadScience00,RoukesPhysWorld01},
such as suspended quantum
dots~\bcite{WeigPRL04}, carbon nanotubes or graphene sheets
~\bcite{LeRoyNature04,BunchScience07}, one-dimensional
wires~\bcite{KriveJLT10}, and molecular
junctions~\bcite{GalperinJoPCM07,ParkNature00}. For these systems, a transport current can
excite mechanical modes, and vice versa, the mechanical motion affects the transport current.The reduced size and high
sensitivity of the resulting devices make them attractive for applications
such as sensors of mass or charge, nanoscale motors, or switches
~\bcite{EomPhysRep11}. On a more
fundamental level, the capability of cooling the system {\it via} back-action
allows one to study quantum phenomena at the mesoscopic level, eventually
reaching the quantum limit of measurement
~\bcite{NaikNature06,StettenheimNature06}.   

All of these applications require an understanding of the mechanical forces that
act on the nanoelectromechanical system in the presence of a transport current. These are referred to as
{\it current-induced forces}, and have been observed
in seminal experiments~\bcite{SteeleScience09,LassagneScience09}. 
Recently we have
shown that it is possible to fully express the current-induced forces in terms
of a scattering
matrix formalism, for arbitrary (albeit adiabatic) out of equilibrium situations
~\bcite{BodePRL11}, thus providing the tools for a systematic approach to study the interplay between electronic and mechanical degrees of freedom in NEMS. 

In
the context of NEMS, two well defined limits can be identified for which
electronic and mechanical time scales decouple, and which give rise to different experimental phenomena. On one side, when the electronic time scales
are slow compared with the mechanical vibrations, drastic consequences can be
observed for the electronic transport, such as side bands due to phonon assisted
tunneling \bcite{YuPRL04,SapmazPRL06} or the Frank-Condon blockade effect, a phononic analog of the
Coulomb blockade in quantum dots \bcite{LeturcqNatPhys09,KochPRL05,KochPRB06}. In
the opposite regime, electrons tunnel
through the nanostructure rapidly, observing a quasistatic configuration of the
vibrational modes, but affecting their dynamics profoundly at the same time
~\bcite{NaikNature06,StettenheimNature06,SteeleScience09,LassagneScience09}. It
is on this regime that our present work focuses. We treat the vibrational
degrees of freedom as classical entities embedded in an electronic environment:
pictorially, many electrons pass through the nanostructure during one
vibrational period, impinging randomly on the modes. In this limit, it is
natural to assume that the dynamics of the vibrational modes, represented by
collective coordinates $X_\nu$, will be governed by a set of coupled Langevin
equations
\begin{equation} 
\label{langevin}
M_\nu \ddot{X}_\nu + \frac{\partial U}{\partial X_\nu} = F_\nu - \sum_{\nu'}
\gamma_{\nu\nu'} \dot{X}_{\nu'} +\xi_\nu\,.
\end{equation}
Here we have grouped the purely elastic contribution on the left hand side (LHS) of
Eq. \eqref{langevin}, $M_\nu$ being the effective mass of mode $\nu$ and
$U({\mathbf X})$ an elastic potential. On the right hand side (RHS) we collected
the current-induced forces: the mean force $F_\nu$, a term proportional to the
velocity of the modes $-\sum_{\nu'}\gamma_{\nu\nu'} \dot{X}_{\nu'}$, and the
Langevin fluctuating forces $\xi_\nu$. The main result of our work are expressions for the current-induced forces in terms of the scattering matrix and its parametric derivatives. These are given by Eq. \eqref{force} for the mean force $F_\nu({\mathbf X})$, Eq. \eqref{variance2} for the correlator $D_{\nu\nu^\prime}({\mathbf X})$ of the stochastic force $\xi_\nu$, and Eqs. (\ref{eq:damp2}) and (\ref{effB2}) for the two kinds of forces (dissipative friction force and effective ``Lorentz'' force, as we discuss below) encoded by the matrix $\gamma_{\nu\nu'}({\mathbf X})$.

Theoretically, these forces have been studied previously within different formalisms. The case of one electronic level coupled to one vibrational mode has been studied with a Green's function approach in Refs. ~\bcite{MozyrskyPRB06,PistolesiPRB08}, where the authors showed that the current-induced forces can lead to a bistable effective potential and consequently to switching. In Ref. \bcite{LueNanoLett10}, the authors studied the case of multiple vibrational modes within a linear approximation, finding a Lorentz-like current-induced force arising from the electronic Berry phase \bcite{BerryPRSL93}. In simple situations, the current-induced forces have been also studied within a scattering matrix approach in the context of quantum measurement backaction \bcite{BennettPRL10} (see also \bcite{BennettPRL10Err}), momentum transfer statistics \bcite{KindermannPRB02}, and of magnetic systems to describe Gilbert damping \bcite{BrataasPRL08}. Current induced forces have been shown to be of relevance near mechanical instabilities \bcite{WeickPRB10,WeickPRB11,MeyerPRB11} and to drive NEMS into instabilities and strong non-linear behavior \bcite{FedoretsEPL02,HusseinPRB10,NoceraPRB11}. Our formalism
allows us to retain the nonlinearities of the problem, which is essential for even
a
qualitative description of the dynamics, while turning the
problem of calculating the current-induced forces into a scattering problem for
which standard techniques can be applied. 

In what follows we develop these ideas in detail, giving a thorough derivation of the expressions in terms of the scattering matrix for the current-induced forces found in Ref. \bcite{BodePRL11}, and include several applications to specific systems. Moreover, we extend the theoretical results of Ref. \bcite{BodePRL11} in two ways. We treat a general coupling between the collective modes $X_\nu$ and the electrons, generalizing the linear coupling expressions obtained previously. We also allow for an arbitrary energy dependence in the hybridization between the leads and the quantum dot, allowing more flexibility for modeling real systems. In Section \ref{sec1} we
introduce the theoretical model, and derive the equations of motion of the mechanical degrees of freedom starting from a microscopic
Hamiltonian. We show how the
Langevin equation, Eq. \eqref{langevin}, emerges naturally from a microscopic model when employing  the
non-equilibrium Born Oppenheimer (NEBO) approximation, appropriate for the limit
of slow vibrational dynamics, and derive the current induced forces in terms of
the microscopic parameters. In Section \ref{sec2} we show that the current induced forces can
be written in terms of parametric derivatives of the scattering matrix
(S-matrix) of the system, and
state general properties that can be derived from S-matrix symmetry
considerations. In Section \ref{sec3} we complete the discussion of nanoelectromechanical systems in terms of scattering matrices by providing a corresponding expression for the charge current. In Section \ref{sec4} we apply our formalism to simple models of increasing complexity, namely a single resonant level, a two-level model, and a two-level/two-mode model. We conclude in Section \ref{sec5}. For better readability, we have relegated part of some lengthy calculations to the Supplementary Material, together with a list of useful relations that are used throughout the main text.

\section{Microscopic derivation of the Langevin equation}
\label{sec1}

\subsection{Model}

We model the system as a mesoscopic quantum dot connected to multiple leads and coupled to vibrational degrees of freedom. Throughout this work we consider non-interacting
electrons and we set $\hbar=1$. The 
Hamiltonian for the full system reads
\beq
H= H_D + H_X + H_L + H_T\,,
\eeq
where the different terms are introduced in the following.

We describe the quantum dot by $M$ electronic levels
coupled to $N$ slow collective degrees of freedom $\hat{{\mathbf X}}=(\hat{X}_1,\ldots,\hat{X}_N)$. This is contained in the dot's Hamiltonian 
\begin{align}
H_D= \sum_{m m'}  d^{\dagger}_{m} \left[h_0(\mathbf{\hat{X}})\right]_{m m'} d_{m'} 
\end{align}
which describes the electronic levels of the dot and their dependence on the collective modes' coordinates $\hat{X}_\nu$ ($\nu=1,\ldots,N$) by the hermitian $M\times M$ matrix
$h_0(\mathbf{\hat{X}})$. The operator $d^{\dagger}$ ($d$) creates (annihilates) an electron in the dot and the indices $m$, $m'$ ($=1,\ldots,M$) label the electronic levels. Note that here we generalize our previous results obtained for a linear coupling in $\mathbf{\hat{X}}$~\bcite{BodePRL11},
and allow $h_0$ to be a general function of $\mathbf{\hat{X}}$. Our analysis is valid for any coupling strength. The free evolution of the `mechanical' degrees of freedom of the dot is described by the 
Hamiltonian
\begin{align}
H_X=  \sum_\nu\left[ \frac{\hat{P}_\nu^2}{2M_\nu}+ U(\hat{\mathbf{X}})\right]\,.
\end{align}

The leads act as electronic reservoirs kept at fixed chemical potentials $\mu_\alpha$ and are described by
\beq
H_{L}=\sum_{\eta} \left(\epsilon_{\eta}-\mu_\alpha\right)
c^{\dagger}_{\eta}c_{\eta}\,,
\eeq
where we represent the electrons in the leads by the creation
(annihilation) operators $c^{\dagger}$ ($c$). The leads' electrons obey the
Fermi-Dirac distribution 
$f_\alpha(\epsilon)=\left[1+e^{(\epsilon-\mu_\alpha)/kT}\right]^{-1}$. The leads are labeled
by $\alpha=1,\ldots,L$, each containing 
channels $n=1,\ldots, N_{\alpha}$.  We combine $\eta=(\alpha,n)$
into a general `lead' index, $\eta=1,\ldots,N_0$ with
$N_0 = \sum_{\alpha} N_{\alpha}$.
  
Finally, the Hamiltonian $H_{T}$ represents the tunneling between the leads and the levels in the
dot,
\beq
  H_{T}= \sum_{\eta,m}  
(c^{\dagger}_{\eta} W_{\eta m} d_{m}+\mathrm{h.c.})\,.
\eeq

\subsection{Non-equilibrium Born-Oppenheimer approximation}
We use as a starting point the Heisenberg equations of motion for the mechanical modes which can be cast as
 \begin{equation}
   M_\nu \ddot{\hat{X}}_\nu + \frac{\partial U}{\partial \hat X_\nu} = - \sum_{n,n^\prime}
d^\dagger_n \left[\Lambda_\nu(\hat {\mathbf X})\right]_{nn^\prime}d_{n^\prime}\, ,
   \label{HEOM} 
\end{equation} 
where we have introduced the $\hat {\mathbf X}$-dependent matrices
\beq\label{LambdaMat}
\Lambda_\nu\left(\hat {\mathbf X}\right) = \frac{\partial h_0}{\partial \hat X_\nu}\,.
\eeq 
The RHS of \eqref{HEOM} contains the current-induced forces, expressed through the electronic operators $d$ of the quantum dot. We now proceed to calculate these forces within a non-equilibrium Born-Oppenheimer (NEBO) approximation, in which the dynamics of the
collective modes is assumed slow. In this limit, we can treat the mechanical degrees of freedom as classical, acting as a slow classical field on the fast electronic dynamics.

The NEBO approximation consists of
averaging the RHS of Eq. \eqref{HEOM} over times long compared to the electronic time scale, but short in terms of the
oscillator dynamics. In this approximation, the force operator is represented by its (average) expectation value $\langle d^\dagger\Lambda d\rangle_{{\mathbf X}(t)}$, evaluated for a given trajectory ${\mathbf X}(t)$ of the mechanical degrees of freedom, plus fluctuations containing both Johnson-Nyquist and shot noise. These fluctuations give rise to a Langevin force $\xi_\nu$. Hence Eq. \eqref{HEOM} becomes
\beq
 M_\nu \ddot X_\nu + \frac{\partial U}{\partial X_\nu} = {\rm tr} [i \Lambda_\nu
{\cal G}^<(t,t)]+\xi_\nu\,,
   \label{NEBO} 
\eeq
where the trace ``${\rm tr}$'' is taken over the dot levels, and we have
introduced
the lesser Green's function 
\beq\label{GlessDef}
{\cal G}_{nn'}^<(t,t')= i\langle
d_{n'}^\dagger(t')d_n^{}(t)\rangle_{{\mathbf X}(t)}\,.
\eeq
The variance of the stochastic force
$\xi_\nu$ is governed by the symmetrized fluctuations of the operator $d^\dagger\Lambda d$. Given that the electronic
fluctuations happen on short time scales, $\xi_\nu$ is locally correlated in time,
\beq\label{defvariance}
\langle \xi_\nu(t) \xi_{\nu'}(t')\rangle=
D_{\nu\nu'}({\bf X}) \delta(t-t')\,.
\eeq
(An alternative but equivalent derivation, is based on a saddle point approximation for the Keldysh action, see {\it e.g.} Ref. \bcite{KamenevAdvPhys09}). Since we are dealing with non-interacting electrons, $D({\bf X})$  can be expressed in terms of single particle Green's functions using Wick's theorem. This readily yields
\beq \label{xixi}
\langle \xi_\nu(t) \xi_{\nu'}(t')\rangle= \mbox{tr}\{\Lambda_\nu {\cal
G}^>(t,t^{\prime}) 
\Lambda_{\nu'} {\cal G}^<(t^{\prime},t)\}_s \,,
\eeq
where
\begin{align}\label{eq:G>}
{\cal G}^{>}_{m m'}(t,t')=-i \langle d_{m}(t)
d_{m'}^{\dagger}(t') \rangle_{{\mathbf X}(t)}\,
\end{align}
is the greater Green's function.
These expressions for the current-induced forces show that we need to evaluate the electronic Green's function for a given classical trajectory ${\mathbf X}(t)$. In doing so, we can exploit that the mechanical degrees of freedom are assumed to be slow compared to the electrons. Thus, we can approximate the Green's function by its solution to first order in the velocities $\dot{\mathbf{X}}(t)$. We now proceed with this derivation, starting with the Dyson equation for the retarded Green's function
\beq\label{GRdef}
{\cal G}^R_{m  m^{\prime}}(t,t^{\prime}) = 
-i \theta(t-t') \langle \{d_{m} (t), d^{\dagger}_{m^{\prime}}(t^{\prime})
\}\rangle_{{\mathbf X}(t)}\,.
\eeq
Here $\{.,.\}$ indicates the anti-commutator. We note that since
we consider non-interacting electrons, we can restore the lesser and greater Green's functions (or the advanced Green's function ${\cal G}^A$) at the end of the calculation by standard manipulations. 

The hybridization with the leads is taken into account through the self-energy \bcite{JauhoPRB94}
\beq\label{SelfEnR}
\Sigma^R(\epsilon)= - i \sum_{\alpha}  \Gamma_\alpha(\epsilon)\,,
\eeq
which is given in terms of the width functions
\beq\label{proj}\Gamma_\alpha(\epsilon)= \pi
W^\dagger(\epsilon) \Pi_\alpha W(\epsilon)\,.
\eeq
Here we have defined  $\Pi_\alpha$ as a projection
operator onto lead
$\alpha$ and absorbed square root factors of the density of states in the leads
into the coupling matrix $W$ for notational simplicity. Note that we allow $W$ to depend on energy. (Compare with the wide-band limit discussed in Ref. \bcite{BodePRL11}, which employs an energy-independent hybridization $\Gamma$.) 

Dyson's
equation for the retarded Green's function can then be written, in matrix form, as
\begin{equation}
-i \partial_{t^{\prime}} {\cal G}^R(t,t^{\prime})= \delta(t-t^{\prime}) 
+  \int dt_1{\cal G}^R(t,t_1)
\Sigma^R(t_1,t^{\prime})+{\cal G}^R(t,t^\prime)h_0({\mathbf X}) 
\,.
\label{dyret}
\end{equation}
To perform the adiabatic expansion, it is convenient to work in the Wigner
representation, in which fast and slow time scales are easily identifiable. The
Wigner
transform of a function $A(t_{1},t_{2})$ depending on two time arguments is
given by  
\begin{equation}
\tilde{A}(t,\epsilon)=\int\mathrm{d}\tau\,e^{i\epsilon\tau}
A(t+\tau/2,
t-\tau/2)\,.
\end{equation}
Using this prescription for
the Green's function ${\cal G}^R$, the slow mechanical motion implies that ${\cal G}^R$ varies slowly
with the central time $t=\frac{t_{1}+t_{2}}{2}$ and oscillates fast with the relative time $\tau=t_{1}-t_{2}$. 
The Wigner transform of a convolution $C(t_{1},t_{2})=\int\mathrm{d}t_{3}\,
A(t_1,t_{3})B(t_{3},t_{2})$
is given by
\begin{eqnarray}\label{moyalexp}
\tilde{C} & = &
\exp\left[\frac{i}{2}\left(\partial_{\epsilon}^{\tilde{A}}\partial_{t}^{\tilde{B
}}
-\partial_ { t } ^{\tilde{A}}\partial_{\epsilon}^{\tilde{B}}\right)\right]\tilde
A\tilde B\nonumber \\
 & \simeq & \tilde A\tilde
B+\frac{i}{2}\partial_{\epsilon}\tilde
A\partial_{t}\tilde
B-\frac{i}{2}\partial_{t}\tilde
A\partial_{\epsilon}\tilde
B,
\end{eqnarray}
where we have dropped higher order derivatives in the last line, exploiting the slow variation with $t$. Therefore, using Eq. \eqref{moyalexp} we can
rewrite the Dyson equation Eq. \eqref{dyret} as
\begin{align}
1\approx {\cal G}^{R}
\left(\epsilon-\Sigma^{R}
-h_0\right)-\frac{i}{2}\partial_{\epsilon}{\cal
G}^{R}\partial_{t}h_0
-\frac{i}{2}\partial_{t}{\cal
G}^{R}\left(1-
\partial_ {\epsilon}\Sigma^{R}\right)\,,
\label{wignerconv}
\end{align}
where the Green's functions are now in the Wigner representation. Unless otherwise denoted by explicitly stating the variables, here and in the following all functions are in the Wigner representation. Finally, with the help of Eqs. \eqref{dgrx} -\eqref{dgre} from Supp. Mat. \ref{app1}, we obtain
\beq
\label{GRadexp}
{\cal G}^R\simeq G^R +\frac{i}{2}\sum_\nu \dot{X}_\nu\left(\partial_\epsilon G^R
\Lambda_\nu G^R - G^R\Lambda_\nu\partial_\epsilon G^R  \right)\,,
\eeq
in terms of the strictly adiabatic Green's function
\beq
\label{FrozenG}
G^{R}(\epsilon,\mathbf{X})=\left[\epsilon-h_{0}(\mathbf{X})
-\Sigma^{R}(\epsilon) \right]^{-1}\,.
\eeq
Our notation is such that ${\cal G}$ denotes {\it full} Green's
functions, while $G$ denotes the strictly adiabatic (or {\it frozen}) Green's functions that are evaluated for a fixed value of ${\mathbf X}$
(so that all derivatives with
respect to central time in Eq. \eqref{wignerconv} can be dropped). From now on, ${\cal G}^{(R,A,<,>)}$ denote the Green functions
in the Wigner representation, with arguments $(\epsilon,t)$, and
${\cal G}^A=({\cal G}^R)^\dagger$. 

Using
Langreth's rule (see {\it e.g.} Ref. \bcite{JauhoPRB94})
\begin{align}\label{eq:Gddless}
  {\cal G}^<(t,t^{\prime})= 
\int\mathrm{d}t_1 \int\mathrm{d}t_2\, {\cal G}^R(t,t_1)
\Sigma^<(t_1,t_2)  {\cal G}^A(t_2,t^{\prime})\,,
\end{align}
we can relate ${\cal G}^<$ with ${\cal G}^R$. In Eq. \eqref{eq:Gddless} we have introduced the lesser self energy $\Sigma^<$, which in the Wigner representation takes the form
\beq\label{SelfEnL}
\Sigma^<(\epsilon) =2
i\, \sum_{\alpha} f_{\alpha}(\epsilon) \Gamma^{\alpha}(\epsilon)\,.
\eeq
Note that $\Sigma^<$ depends only on $\epsilon$ and is independent of the central time. Expanding Eq. \eqref{eq:Gddless} up to the leading adiabatic correction according to Eq. \eqref{moyalexp}, we obtain ${\cal G}^<$ to first order in $\dot{{\mathbf X}}$, 
\begin{equation}\label{GLadexp}
{\cal G}^< = G^< +\frac{i}{2} \sum_\nu\dot X_\nu 
\left[ (\partial_\epsilon G^<) \Lambda_\nu G^A - 
G^R \Lambda_\nu \partial_\epsilon G^< + 
(\partial_\epsilon G^R) \Lambda_\nu G^< 
- G^< \Lambda_\nu \partial_\epsilon G^A \right ]\,, 
\end{equation}
with $G^<=G^R\Sigma^<G^A$.
 
\subsection{Current-induced forces in terms of Green's functions}
We can now collect the results from the previous section and identify the current-induced forces appearing in the Langevin equation \eqref{langevin}. Except for the stochastic noise force, the
current induced 
forces are encoded in ${\rm tr}({\cal G}^<\Lambda_\nu)$. In the strictly
adiabatic limit, {\it i.e.}, retaining only the first term on the RHS of Eq.~(\ref{GLadexp}), ${\cal G}^<
\simeq G^<$, we obtain the mean force
\beq\label{fdef}
F_\nu({\mathbf X}) = - \int \frac{d\epsilon}{2\pi i}  \ {\rm
tr}\left[\Lambda_\nu G^<  \right]\,.
\eeq

The leading order correction in 
Eq.~(\ref{GLadexp}) gives a velocity-dependent contribution to the current induced forces, which determines the tensor $\gamma_{\nu\nu'}$. After integration by parts, we find
\beq\nonumber
\gamma_{\nu\nu'}=\int\frac{d\epsilon}{2\pi}
{\rm tr} \left( G^<\Lambda_\nu \partial_\epsilon G^R \Lambda_{\nu'} 
- G^< \Lambda_{\nu'} \partial_\epsilon G^A  \Lambda_\nu  \right).
\eeq
This tensor can be split into symmetric and  anti-symmetric contributions, $\gamma
= \gamma^s + \gamma^a$, which define a dissipative term $\gamma^s$ and an
orbital, effective magnetic field $\gamma^a$ in the space of the collective modes. The latter interpretation is based on the fact that the corresponding force takes a Lorentz-like form. Using Eq. \eqref{GRGAG>G<} in the Supp. Mat. \ref{app1} and noting that $2 \int d\epsilon G^<
\partial_\epsilon G^< = \int d\epsilon \partial_\epsilon(G^<)^2=0$, 
we obtain the explicit expressions
\begin{widetext}
\begin{eqnarray} \label{sdef}
\gamma^s_{\nu\nu'}({\mathbf X}) &=& \int \frac{d\epsilon}{2\pi} {\rm tr}
\left \{ \Lambda_\nu G^< \Lambda_{\nu'}
 \partial_\epsilon G^> \right \}_s ,\\ \label{adef}
\gamma^a_{\nu\nu'}({\mathbf X}) &=& -\int \frac{d\epsilon}{2\pi} {\rm tr}
\left \{ \Lambda_\nu G^< \Lambda_{\nu'}
 \partial_\epsilon \left(G^A+G^R\right) \right \}_a \,.
\end{eqnarray}
\end{widetext}
Here we have introduced the notation 
\[
\{ A_{\nu\nu'} \}_{s,a} = \frac12 (A_{\nu\nu'}\pm A_{\nu'\nu} )\,
\]
for symmetric and anti-symmetric parts of an arbitrary matrix $A$.

At last, the stochastic force
$\xi_\nu$ is given by the thermal and non-equilibrium fluctuations of the force
operator $-d^\dagger \Lambda_\nu d$ in Eq. \eqref{HEOM}. As indicated by the fluctuation-dissipation theorem, the fluctuating force is of the same order in the adiabatic expansion as the velocity dependent force. Thus, we can evaluate the expression for the correlator $D_{\nu\nu'}({\mathbf X})$ of the fluctuating force given in  Eq. \eqref{xixi} to lowest order in the adiabatic expansion, so that 
\beq \label{ddef}
D_{\nu\nu'}({\mathbf X}) = \int \frac{d\epsilon}{2\pi} {\rm tr}
\left\{ \Lambda_\nu G^< \Lambda_{\nu'}
G^> \right\}_s.
\eeq
This formalism gives the tools
needed to describe the dynamics of the vibrational modes in the presence of a
bias for an arbitrary number of modes and dot levels.
When expressions \eqref{fdef} - \eqref{adef} are inserted back in Eq.
\eqref{langevin}, they define a non-linear Langevin equation due to their non-trivial dependences on $\mathbf{X}(t)$ ~\bcite{MozyrskyPRB06,PistolesiPRB08}.  

\section{S-matrix theory of current-induced forces}
\label{sec2}
\subsection{Adiabatic expansion of the S-matrix}
Scattering matrix approaches to mesoscopic transport generally
involve expressions in terms of the elastic S-matrix. For our problem, the S-matrix is elastic only in the strictly adiabatic limit, in which it is evaluated for a fixed value of ${\bf X}$,
\begin{equation}
\label{FrozenS}
  S(\epsilon,{\bf X}) = 1 - 2\pi i W(\epsilon) G^R(\epsilon,{\bf X}) W^\dagger(\epsilon)\,.
\end{equation}
As pointed out by Moskalets and B\"uttiker~\bcite{MoskaletsPRB04,MoskaletsPRB05},
this is not sufficient for general out of equilibrium
situations, even when ${\bf X}(t)$ varies in time adiabatically. In their work, they calculated, within a Floquet formalism, the
leading correction to the strictly adiabatic S-matrix. We follow here the same approach, rephrased in terms of the Wigner representation. The full S-matrix can be written as \bcite{AleinerPhysRep02} (note
that, in line with the notation established before for the Green's functions, the strictly adiabatic S-matrix
is denoted by $S$ while the full S-matrix is denoted by ${\cal S}$)
\beq
\label{FullS}
 {\cal S}(\epsilon,t)= 1-2\pi i \left[W {\cal G}^R W^\dagger\right](\epsilon,t)\,.
\eeq
To go beyond the frozen approximation, we expand ${\mathcal S}$ to leading order in $\dot{\mathbf X}$,
\beq
\label{Sadexp}
{\cal S}(\epsilon,t) \simeq S(\epsilon,{\bf X}(t))+ \sum_\nu \dot X_\nu(t)
A_\nu(\epsilon,{\bf X}(t))\,.
\eeq
Thus, the leading correction defines the matrix $A$, which, similar to $S$, has definite symmetry properties. In particular, if the system is time-reversal invariant, the adiabatic S-matrix is even under time reversal while $A$ is odd. For a given problem, the A-matrix has to
be obtained along with $S$.

We can now derive a Green's function expression for the matrix $A$~\bcite{VavilovPRB01,ArracheaPRB06}. Comparing Eq. \eqref{Sadexp} with the expansion to the same order of ${\cal S}$ in
terms of adiabatic Green's functions (obtained straightforwardly by performing explicitly the convolution in Eq. \eqref{FullS} and keeping terms up to $\dot
{\mathbf X}$) we obtain
\beq
\begin{split}
\label{Amatrix}
A_\nu(\epsilon,{\bf X}) &= \pi  \partial_\epsilon \left[W(\epsilon) G^R(\epsilon,{\bf X})\right] \Lambda_\nu({\bf X}) G^R(\epsilon,{\bf X})
W^\dagger(\epsilon)\\&- \pi
W(\epsilon) G^R(\epsilon,{\bf X}) \Lambda_\nu ({\bf X})\partial_\epsilon \left[G^R(\epsilon,{\bf X}) W^\dagger(\epsilon)\right]\,.
\end{split}
\eeq
Current conservation constrains both the frozen and full scattering matrices to be unitary.
From the unitarity of the frozen S-matrix, $S^\dagger S={\mathbf 1}$, we obtain the useful relation
\beq\label{UnFrozS}
\frac{\partial S^\dagger} {\partial X_\nu} S +
S^\dagger\frac{\partial S} {\partial X_\nu} =0\,.
\eeq
We will make use of Eq. \eqref{UnFrozS} repeatedly in the following sections. 
On the other hand, unitarity of the full S-matrix, ${\cal S}^\dagger{\cal
S}={\mathbf 1}$, imposes a relation between the A-matrix and the frozen S-matrix. To first order in the velocity $\dot{\mathbf X}$ we have
\begin{align}
{\mathbf 1}= S S^\dagger + S A^\dagger + A S^\dagger
+\frac{i}{2} \left(\frac{\partial S}{\partial \epsilon} \frac{\partial
S^\dagger}{\partial t} - \frac{\partial S}{\partial t} \frac{\partial
S^\dagger}{\partial \epsilon}\right)
\end{align}
where $ A(\epsilon,\mathbf X) = \sum_{\nu} A_{\nu}(\epsilon,\mathbf X)
\dot{X}_{\nu}$. Therefore, $S$ and $A$ are related through
\begin{align}\label{eq:ASdag}
  A_{\nu} S^{\dagger} + S A_{\nu}^{\dagger} = \frac{i}{2} \left(\frac{\partial
S}{\partial X_{\nu}} \frac{\partial
S^{\dagger}}{\partial \epsilon} - \frac{\partial
S}{\partial \epsilon} \frac{\partial
S^{\dagger}}{\partial X_{\nu}}  \right)\,.
\end{align}

In the next section we will see that the A-matrix is essential to express the
current-induced dissipation and ``Lorentz'' forces, Eqs. \eqref{sdef} and
\eqref{adef}.

\subsection{Current-induced forces}  
\subsubsection{Mean Force}
The mean force exerted by the electrons on the oscillator is given by Eq.
\eqref{fdef}. Writing Eq. \eqref{fdef} explicitly and using Eq. \eqref{eq:G<ad} in Supp. Mat. \ref{app1}, we can express $G^<$ in terms of $G^R$ and $G^A$ and obtain
\beq\label{auxF1}
\begin{split}
F_\nu({\mathbf X}) &= - \int d\epsilon \sum_\alpha f_\alpha {\rm
tr}\left(\Lambda_\nu G^R  W^\dagger\Pi_\alpha W G^A
 \right)\,\\
&= - \int d\epsilon \sum_\alpha f_\alpha {\rm
tr}\left(\Pi_\alpha W G^A \Lambda_\nu G^R  W^\dagger\right)\,,
\end{split}
\eeq 
where the second equality exploits the cyclic
invariance of the 
trace. Noting that, by Eq. \eqref{sder} in Supp. Mat. \ref{app1},
\begin{equation} 
W G^A \Lambda_\nu G^R  W^\dagger=-\frac{1}{2\pi i }S^\dagger \frac{\partial S}{\partial X_{\nu}}\,,
\end{equation}
Eq. \eqref{auxF1} can be expressed directly in
terms of scattering matrices $S(\epsilon,{\bf X})$ as
\begin{equation}
\label{force}
F_\nu ({\mathbf X})=  \sum_\alpha\int \frac{d\epsilon}{2\pi i} f_\alpha
{\rm Tr} \left( \Pi_\alpha S^\dagger \frac{\partial S}{\partial X_\nu} 
\right). 
\end{equation}
Note that now the trace (denoted by ``${\rm Tr}$'') is over lead-space. 

An important issue is whether this force is \textit{conservative}, 
{\it i.e.}, derivable from a potential.  A necessary condition for this is a vanishing  
``curl'' of the force,
\begin{eqnarray}\label{curl}
\Omega_{\nu \nu'} \equiv\frac{\partial F_{\nu'}}{\partial X_{\nu}}-\frac{\partial
F_{\nu}}
{\partial X_{\nu'}} 
= \sum_\alpha \int \frac{d\epsilon}{\pi i}
f_\alpha  \ {\rm Tr} 
\left( \Pi_\alpha \frac{\partial S^\dagger}{\partial X_\nu}
\frac{\partial S}{\partial X_{\nu'}} \right)_a\,.
\end{eqnarray}
From Eq. \eqref{curl} it is seen that the mean force is conservative in thermal equilibrium, where Eq. \eqref{curl} can
be
turned into a trace over a commutator of finite-dimensional matrices: Indeed, in
equilibrium the sum over the lead indices can be directly performed since
$f_\alpha=f$ for all $\alpha$, and $\sum_\alpha \Pi_\alpha=1$. Using the unitarity of the S-matrix and
the cyclic property of the trace, we obtain:
\beq
\begin{split}\label{curleq}
\Omega_{\nu \nu'}  &= \int \frac{d\epsilon}{2 \pi i}
f  \ {\rm Tr} 
\left(\frac{\partial S^\dagger}{\partial X_\nu}
\frac{\partial S}{\partial X_{\nu'}} - \frac{\partial S^\dagger}{\partial
X_{\nu'}} S S^\dagger
\frac{\partial S}{\partial X_{\nu}} \right)\\
  &= \int \frac{d\epsilon}{2 \pi i}
f  \ {\rm Tr} 
\left(\frac{\partial S^\dagger}{\partial X_\nu}
\frac{\partial S}{\partial X_{\nu'}} -\frac{\partial S}{\partial
X_{\nu'}}
\frac{\partial S^\dagger}{\partial X_{\nu}} \right) =0\,,
\end{split}
\eeq
where in the last line we have used Eq. \eqref{UnFrozS}. In general, however, the
mean force will be {\em non-conservative} in out-of-equilibrium situations, providing a way to exert work on the mechanical degrees of freedom by controlling the external bias potential~\bcite{DundasNatNano09,TodorovPRB10,LueNanoLett10}.

\subsubsection{Stochastic Force}
Next, we discuss the fluctuating
force $\xi_\nu$ with variance $D_{\nu\nu'}$ given by Eq. \eqref{ddef}. Following
a similar path as described in the previous subsection for the mean
force $F_\nu$ , we can also express the variance Eq.~(\ref{ddef})
of the fluctuating force in terms of the adiabatic S-matrix,
\beq
\label{variance2}
D_{\nu\nu'}({\bf X})= \sum_{\alpha\alpha'} \int\frac{d\epsilon}{2\pi}F_{\alpha \alpha'}
{\rm Tr} \left\{ \Pi_\alpha \left[ S^\dagger 
\frac{\partial S}{\partial X_{\nu}}\right]^\dagger \Pi_{\alpha'} S^\dagger 
\frac{\partial S}{\partial X_{\nu'}} \right\}_s\,,
\eeq
where we have introduced the function $F_{\alpha \alpha'}(\epsilon)=f_\alpha(\epsilon) \left[1-f_{\alpha'}(\epsilon) \right]$. From Eq. \eqref{variance2} it is straightforward to show that $D_{\nu\nu'}$ is positive definite.
By performing a unitary transformation to a basis in which $D_{\nu\nu'}$ is
diagonal, using
$\Pi_\alpha=\Pi_\alpha^2$ and the cyclic invariance of the trace, we obtain the expression 
\beq\label{Dpos}
D_{\nu\nu}({\bf X})= \sum_{\alpha\alpha'} \int\frac{d\epsilon}{2\pi}F_{\alpha \alpha'}{\rm Tr} \left\{ \left(\Pi_{\alpha'}  S^\dagger 
\frac{\partial S}{\partial X_{\nu}}\Pi_\alpha\right)^\dagger \Pi_{\alpha'} S^\dagger 
\frac{\partial S}{\partial X_{\nu}}\Pi_\alpha \right\}_s\,.
\eeq
which is evidently positive.

\subsubsection{Damping Matrix}
So far, we were able to express quantities in terms of the frozen S-matrix only. This is no longer the case for the first correction to the
strictly adiabatic approximation, given by Eqs. \eqref{sdef} and \eqref{adef}. We start here with the first of these
terms, the symmetric matrix $\gamma^s$, which is responsible for dissipation of
the mechanical system into the electronic bath.

The manipulations to write the dissipation term as a function of S-matrix quantities are lengthy and the details are given in the Supp. Mat. \ref{app2}. The damping matrix can be split into an ``equilibrium'' contribution,
$\gamma^{s,eq}$, and a purely non-equilibrium contribution $\gamma^{s,ne}$, as 
$\gamma^{s}=\gamma^{s,eq}+\gamma^{s,ne}$.
We first treat $\gamma^{s,eq}$. By the calculations given in Supp. Mat. \ref{app2}, we obtain
\beq
\begin{split}
\label{gseq}
  \gamma^{s,eq}_{\nu\nu'}&=\frac14 \sum_{\alpha\alpha'} \int \frac{d\epsilon}{2\pi}
\partial_\epsilon( f_\alpha + f_{\alpha'}) {\rm Tr} \left\{
\Pi_\alpha S^\dagger \frac{\partial S}{\partial X_\nu}
 \Pi_{\alpha'} S^\dagger \frac{\partial S}{\partial X_{\nu'}} \right\}_s \\
&=\frac12 \sum_{\alpha} \int \frac{d\epsilon}{2\pi}
(-\partial_\epsilon f_\alpha) {\rm Tr} \left(
\Pi_\alpha \frac{\partial S^\dagger}{\partial X_\nu}
 \frac{\partial S}{\partial X_{\nu'}} \right) \,,
\end{split}
\eeq
where we have used that $\sum_{\alpha'}\Pi_{\alpha'}=1$ , $S^\dagger S=1$, and Eq. \eqref{UnFrozS} in the last line. Note that in general, $\gamma^{s,eq}$
also contains non-equilibrium
contributions, but gives the only contribution to the damping matrix when in
equilibrium. Eq. \eqref{gseq} is analogous to the S-matrix expression obtained for dissipation in ferromagnets in thermal equilibrium, dubbed Gilbert damping~\bcite{BrataasPRL08}. 

To express $\gamma^{s,ne}$ in terms of S-matrix quantities, we have to
make use of the A-matrix defined in Eq. \eqref{Amatrix}. Again the details are given in  the Supp. Mat. \ref{app2}, where we find after lengthy manipulations that
\beq\label{gammaneqS}
  \gamma^{s,ne}_{\nu\nu'}= \int \frac{d\epsilon}{2\pi i}
\sum_\alpha f_\alpha {\rm Tr}\left\{ \Pi_\alpha \left(\frac{\partial
S^\dagger}{\partial X_\nu} 
A_{\nu'} - A^\dagger_{\nu'} \frac{\partial S}{\partial X_{\nu}} \right)
\right\}_s.
\eeq
This quantity vanishes in equilibrium, as can be shown using the properties of
the $S$ and $A$ matrices. Since the sum over leads can be directly performed in equilibrium, expression \eqref{gammaneqS} involves
\begin{align}
  {\rm Tr} \left\{\frac{\partial S^\dagger}{\partial X_\nu} 
A_{\nu'} - A^\dagger_{\nu'} \frac{\partial S}{\partial X_{\nu}} \right
\}_s=&-{\rm Tr} \left\{\frac{\partial S}{\partial X_\nu} S^{\dagger}
\left(A_{\nu'} S^\dagger + S A_{\nu'}^\dagger \right) \right \}_s \nonumber\\
=&- \frac{i}{2} {\rm Tr} \left\{\frac{\partial S}{\partial X_\nu} S^{\dagger}
\left(\frac{\partial
S}{\partial X_{\nu'}} \frac{\partial
S^{\dagger}}{\partial \epsilon} - S \frac{\partial
S^\dagger}{\partial \epsilon} \frac{\partial
S}{\partial X_{\nu'}}  S^\dagger  \right) \right \}_s
  =0
\end{align}
where we have used the unitarity of ${\cal S}$ and the cyclic invariance of
the trace multiple times.
In the first equality, we inserted $S^\dagger S=1$ and used Eq. \eqref{UnFrozS}, the
second equality follows by inserting the identity \eqref{eq:ASdag} and using
again \eqref{UnFrozS}.

Finally, combining all terms we obtain an S-matrix expression for the full damping matrix $\gamma^s$, 
\beq\label{eq:damp2}
\begin{split}
  \gamma^s_{\nu\nu'} ({\bf X}) &= -\sum_\alpha \int \frac{d\epsilon}{4\pi}
\partial_\epsilon
f_\alpha {\rm Tr} \left\{ \Pi_\alpha \frac{\partial S^\dagger} {\partial X_\nu}
\frac{\partial S}{\partial X_{\nu'}}\right\}_s
\\&+ \sum_{\alpha} \int
\frac{d\epsilon}{2\pi i} f_\alpha
{\rm Tr} \left\{ \Pi_{\alpha}\left(\frac{\partial S^\dagger}{\partial X_\nu} 
A_{\nu'} - A^\dagger_{\nu'} \frac{\partial S}{\partial X_{\nu}} \right)  \right
\}_s.
\end{split}
\eeq

Note that in equilibrium, by the relation $-\partial_\epsilon f=f(1-f)/T$ and using Eq. \eqref{UnFrozS}, the fluctuating force $D$ and damping $\gamma^s$ are related via
\begin{align}
  D_{\nu \nu'} = 2 T \gamma^{s,eq}_{\nu \nu'}=2 T \gamma^s_{\nu \nu'}\,
\end{align}
as required by the fluctuation-dissipation theorem.

Following a similar set of steps as shown above for the variance $D_{\nu \nu'}$ in Eq. \eqref{Dpos}, $\gamma^{s,eq}_{\nu \nu'}$ has positive eigenvalues. On the other hand,
the sign of $\gamma^{s,ne}_{\nu \nu'}$ is
not fixed, allowing the possibility of negative eigenvalues of $\gamma^s$. The
possibility of negative damping is, therefore, a pure non-equilibrium effect. Several recent papers found negative damping in specific out
of equilibrium models ~\bcite{ClerkNJP05,HusseinPRB10,BodePRL11,LuePRL11}. 

\subsubsection{Lorentz force}
We turn now to the remaining term, the antisymmetric contribution $\gamma^a$
given in Eq.~(\ref{adef}), which acts as an effective magnetic field. Using Eq.
\eqref{eq:G<ad} in Supp. Mat. \ref{app1}, it can be written as
\begin{equation}
\gamma^a_{\nu\nu'} = i \int d\epsilon\sum_\alpha f_\alpha
{\rm Tr}\left\{\Pi_\alpha W G^A 
\Lambda_\nu \left(\partial_\epsilon G^R+\partial_\epsilon G^A\right) \Lambda_{\nu'} G^R
W^\dagger \right\}_a.
\end{equation} 
In order to relate this to the scattering matrix, we use the Supp. Mat. \ref{app1} Eq. \eqref{sdagdxa},
which allows us to write $\gamma^a$ in terms of the
S-matrix as
\begin{equation}
\gamma^a_{\nu\nu^\prime}({\bf X}) = \sum_{\alpha} \int \frac{d\epsilon}{2\pi i}
f_\alpha
{\rm Tr} \left\{ \Pi_{\alpha}\left( S^\dagger \frac{\partial A_{\nu}}{\partial
X_{\nu'}} -
\frac{\partial A^\dagger_{\nu}}{\partial X_{\nu'}} S\right)  \right \}_a \,.
\label{effB2}
\end{equation}

If the system is time-reversal invariant, $\gamma^a$ vanishes in thermal
equilibrium. The latter implies $
\sum_\alpha \Pi_\alpha f_\alpha = f$, so that Eq. \eqref{effB2} involves only
\begin{align}\nonumber
  \mathrm{Tr}\left\{ S^\dagger \frac{\partial A_{\nu}}{\partial
X_{\nu'}} -
\frac{\partial A^\dagger_{\nu}}{\partial X_{\nu'}} S \right \} &
    =\mathrm{Tr}\left\{\frac{\partial A^T_{\nu}}{\partial
X_{\nu'}} S^* - S^T \frac{\partial A^*_{\nu}}{\partial X_{\nu'}} \right
\}=\mathrm{Tr}\left\{-\frac{\partial A_{\nu}}{\partial
X_{\nu'}} S^\dagger + S \frac{\partial A^\dagger_{\nu}}{\partial X_{\nu'}}
\right \},
\end{align}
yielding $\gamma^a=0$ due to the cyclic invariance of the trace. In the last
equality, we
have used $S=S^T$ and $A=-A^T$ as implied by time-reversal invariance.

Out of equilibrium, $\gamma^a$ generally does not
vanish even for time reversal symmetric conductors, since the current effectively breaks
time reversal symmetry.

\section{Current}
\label{sec3}
So far we have focused on the effect of the electrons on the mechanical degrees of freedom. For a complete picture, we also need to consider the reverse effect of the mechanical vibrations on the electronic current. In the strictly adiabatic limit, this obviously has to reduce to the Landauer-B\"uttiker formula for the transport current. Considering the leading adiabatic correction to the current in equilibrium is closely related to the phenomenon of quantum pumping, and we will see that our results in this limit essentially reduce to Brouwer's S-matrix formula for the pumping current \bcite{BrouwerPRBR98}. Our full result is, however, more general since it gives the leading adiabatic correction to the current in arbitrary {\it non-equilibrium} situations \bcite{MoskaletsPRB05}.

The current through lead $\alpha$ is given by
~\bcite{JauhoPRB94}:
\begin{align}
  I_{\alpha} &=-e \langle \dot{N}_{\alpha}\rangle =i e \sum_{n, \eta
\in \alpha} W_{\eta n} \langle c_\eta^\dagger(t) d_n(t) \rangle+
\mathrm{h.c.}
\end{align}
with $N_{\alpha} = \sum_{\eta\in \alpha} c_{\eta}^{\dagger} c_{\eta}$. Using the
expressions for the
self-energies this can be expressed in terms of the dot's Green's functions and
self-energies,
\beq
  I_{\alpha}(t) =e\int \mathrm{d}t^{\prime}\, \mathrm{tr}\left\{
{\cal G}^R(t,t^{\prime})
\Sigma_{\alpha}^<(t^{\prime},t)+ {\cal G}^<(t,t^{\prime})
\Sigma_{\alpha}^A(t^{\prime},t)\right\} + \mathrm{h.c.}\,.
\eeq
Again we use the separation of time scales and go to the Wigner representation, yielding
\beq\label{eq:IWigner}
 I_{\alpha} = e \int \frac{\mathrm{d}\epsilon}{2\pi}\,
\mathrm{tr}\left\{{\cal G}^R \Sigma^<_{\alpha} + {\cal
G}^< \Sigma^A_{\alpha} -\tfrac{i}{2}\left( \partial_t
{\cal G}^R \partial_{\epsilon} \Sigma_{\alpha}^<+\partial_t
{\cal G}^< \partial_{\epsilon} \Sigma_{\alpha}^A\right)\right\}
+\mathrm{h.c.}.
\eeq
We split the current into an adiabatic contribution $I^0_{\alpha}$ and a term
proportional to
the velocity $\dot{X}_{\mu}$: 
\begin{align}
  I_{\alpha} = I^0_{\alpha} + I^{1}_{\alpha}. 
\end{align} 
We will express these quantities in terms of the scattering matrix. 

\subsection{Landauer-B\"uttiker current}
The strictly adiabatic contribution to the current is given by
\begin{align}
  I^0_{\alpha}({\bf X}) &=e \int \frac{\mathrm{d}\epsilon}{2\pi}\,
\mathrm{tr}\left\{
\left({G}^R-{G}^A\right)
\Sigma_{\alpha}^<+ {G}^<
\left(\Sigma_{\alpha}^A-\Sigma_{\alpha}^R\right)\right\}\,,
\end{align}
where we have collected the purely adiabatic terms from Eqs. \eqref{GRadexp} and \eqref{GLadexp}. Inserting the expressions for the self-energies Eqs. \eqref{SelfEnR} and \eqref{SelfEnL}, we can express this as
\beq
 I^0_{\alpha}({\bf X})= e \int \frac{\mathrm{d}\epsilon}{2\pi}\,
\sum_{\beta} f_{\beta} 2\pi i \,
\mathrm{Tr}\left\{W \bigl[\delta_{\alpha
\beta} (G^R-G^A)+ 2\pi i G^R W^{\dagger}
\Pi_{\beta} W G^A \bigr]W^{\dagger} \Pi_{\alpha}\right\},
\eeq
where we used Supp. Mat. \ref{app1} Eq. \eqref{eq:G<ad}. Inserting the adiabatic S-matrix, Eq. \eqref{FrozenS} yields
\begin{align}
I^0_{\alpha}({\bf X})&= e \int \frac{\mathrm{d}\epsilon}{2\pi}\,
\sum_{\beta} f_{\beta} \mathrm{Tr}\left\{\left[\delta_{\alpha \beta} -
S  \Pi_{\beta} S^{\dagger} \right] \Pi_{\alpha}\right\}\\
  &= e \int \frac{\mathrm{d}\epsilon}{2\pi}\,
\sum_{\beta} \left(f_{\alpha}-f_{\beta}\right)\,
\mathrm{Tr}\left\{
S  \Pi_{\beta} S^{\dagger}  \Pi_{\alpha}\right\}\,,
\end{align}
where we used $\sum_\beta S \Pi_\beta S^\dagger =1$ in the last line. We hence recover the usual expression for the Landauer-B\"uttiker current~\bcite{ButtikerPRB85}. Note that the total adiabatic current depends implicitly on time through ${\bf X}(t)$, and is conserved at every instant of time, $\sum_{\alpha}I^0_{\alpha}({\bf X}) = 0$. To obtain the {\em dc} current, we need to average this expression over the Langevin dynamics of the mechanical degrees of freedom. Alternatively, we can average the current expression with the probability distribution of ${\bf X}$, which can be obtained from the corresponding Fokker-Planck equation. Similar remarks would apply to calculations of the current noise.

\subsection{First order correction}
Now we turn to the first order correction to the adiabatic approximation \bcite{MoskaletsPRB05}, restricting our considerations to the wide-band limit. The contribution to the current \eqref{eq:IWigner} which is linear in the velocity reads
\begin{widetext}
\beq
  I^1_{\alpha}({\bf X})=e \int \frac{\mathrm{d}\epsilon}{2\pi}\,
i \sum_{\mu}\dot{X}_{\mu}
\,\mathrm{tr}\left\{(\partial_{\epsilon}G^R)\Lambda_{\mu}
G^R \Sigma_{\alpha}^< + \left[\left(\partial_{\epsilon}G^<\right)\Lambda_{\mu} G^A- G^R
\Lambda_{\mu}(\partial_{\epsilon}G^<)\right]\Sigma_{\alpha}^A\right\}+\mathrm{
h.c.},
\eeq
\end{widetext}
after integration by parts.
Again, we insert Eq. \eqref{eq:G<ad} from Supp. Mat. \ref{app1} for the lesser Green's function, and expressions \eqref{SelfEnR} and \eqref{SelfEnL} for the self-energies.
In the wide band limit, the identity $(i/2) \partial_\epsilon \partial_{X_\nu} S + A_\nu = W (\partial_\epsilon G^R) \Lambda_\nu G^R W^\dagger$ holds, so that we can write
\beq
I^1_{\alpha}({\bf X})= - e \int \frac{\mathrm{d}\epsilon}{2\pi}\,
\dot{\mathbf{X}} \cdot\sum_{\beta} f_{\beta}\mathrm{Tr}\left[\left(\frac{i}{2} \frac{\partial^2
S}{\partial \epsilon \partial \mathbf{X}} +
{\bf A}\right)\Pi_{\beta}S^{\dagger}\Pi_{\alpha}\right] +\mathrm{h.c.}\,
\eeq
after straightforward calculation. After integration by parts, we can split this expression as 
\begin{widetext}
\beq
\begin{split}
I^1_{\alpha}({\bf X}) &= - \frac{e}{2\pi} \int d\epsilon 
 \dot{\bf X}\cdot\sum_\beta \partial_\epsilon f_\beta 
 {\rm Im}{\rm Tr} \left\{ \Pi_\alpha \frac{\partial S}{\partial {\bf X}}
\Pi_\beta S^\dagger \right\}\\
& + \frac{e}{2\pi} \int d\epsilon\dot{\bf X}\cdot \sum_\beta f_\beta \ {\rm Re}{\rm Tr} \left\{ i\Pi_\alpha \frac{\partial
S}{\partial{\bf X}} \Pi_\beta \frac{\partial S^\dagger}{\partial \epsilon} - 2\Pi_\alpha {\bf A} \Pi_\beta S^\dagger  \right\} .\label{pump}
\end{split}
\eeq
\end{widetext}
In equilibrium, the second term vanishes due to the identity Eq. \eqref{eq:ASdag} and the first term agrees with Brouwer's formula for the pumping current~\bcite{BrouwerPRBR98}. As for the strictly adiabatic contribution, the {\em dc} current is obtained by averaging over the probability distribution of ${\bf X}$.

\section{Applications}
\label{sec4}
\subsection{Resonant Level}
To connect with the existing literature, as a first example we treat the simplest case within our formalism: a resonant level coupled to a single vibrational mode and attached to two leads on the left ($L$) and right ($R$). This model has been discussed in detail for zero temperature in references \bcite{MozyrskyPRB06,PistolesiPRB08}, and it provides a simple description on how current-induced forces can be used to manipulate a molecular switch. Here we derive finite-temperature expressions for the current-induced forces for a generic coupling between electronic and mechanical degrees of freedom, starting from the scattering matrix of the system, and show how they reduce to the known results for zero temperature and linear coupling.

We consider  $N=M=1$, denoting the mode coordinate by $X$, the energy of the dot level by $\tilde \epsilon(X)$, and the number of channels in the left and right leads by $N_L$ and $N_R$, respectively. The Hamiltonian of the dot can then be written as 
\beq
H_D=\tilde\epsilon(X)d^\dagger d
\eeq
and the hybridization matrix as $W^\dagger=\left({\mathbf w}^L,{\mathbf w}^R\right)^\dagger$, with ${\mathbf w}^\alpha=(w^\alpha_1,\ldots w^\alpha_{N_\alpha})$ and $\alpha=L,\,R$. Hence the frozen S-matrix, Eq. \eqref{FrozenS}, is given by 
\beq\label{SmatResLev0}
S= {\mathbf 1}-\frac{2\pi i}{{\mathcal L}} \left
( \begin{array}{cc} {\mathbf w}^L\left({\mathbf w}^L\right)^\dagger & {\mathbf w}^L\left({\mathbf w}^R\right)^\dagger\\ 
{\mathbf w}^R\left({\mathbf w}^L\right)^\dagger& {\mathbf w}^R\left({\mathbf w}^R\right)^\dagger \end{array} \right)\,,
\eeq
where ${\mathcal L}(\epsilon, X)=\epsilon-\tilde\epsilon(X)+i\Gamma$, $\Gamma=\Gamma_L+\Gamma_R$, and $\Gamma_\alpha=\pi\left({\mathbf w}^\alpha\right)^\dagger\cdot{\mathbf w}^\alpha$. Rotating to an eigenbasis of the lead channels, this S-matrix does not mix channels within the same lead, and hence we can project the S-matrix into a single non-trivial channel in each lead, to obtain
\beq\label{SmatResLev}
S={\mathbf 1}-\frac{2i}{{\mathcal L}} \left
( \begin{array}{cc} \Gamma_L & \sqrt{\Gamma_L\Gamma_R} 
 \\ \sqrt{\Gamma_L\Gamma_R} & \Gamma_R \end{array} \right)\,.
\eeq
To calculate the mean force from Eq. \eqref{force}, we need an explicit expression for Eq. \eqref{sder} in Supp. Mat. \ref{app1}. This can be easily calculated to be 
\beq
S^\dagger\frac{\partial S}{\partial X}=- \frac{\partial \tilde \epsilon}{\partial X}\frac{2 i}{\left|{\mathcal L}\right|^2} \left
( \begin{array}{cc} \Gamma_L & \sqrt{\Gamma_L\Gamma_R} 
 \\ \sqrt{\Gamma_L\Gamma_R} & \Gamma_R \end{array} \right)\
\eeq 
and hence 
\beq
F(X)=-\int\frac{d\epsilon}{\pi}\left[\frac{f_L\Gamma_L+f_R\Gamma_R}{\left|{\mathcal L}\right|^2}\right]\frac{\partial \tilde \epsilon}{\partial X}\,.
\eeq
Analogously, the variance of the stochastic force, Eq. \eqref{variance2}, becomes
\beq
D(X)=2\int\frac{d\epsilon}{\pi}\sum_{\alpha\alpha'}\frac{\Gamma_\alpha\Gamma_{\alpha'}F_{\alpha\alpha'}}{\left|{\mathcal L}\right|^4}\left[\frac{\partial \tilde \epsilon}{\partial X}\right]^2\,.
\eeq
It only remains to calculate the dissipation coefficient $\gamma$. Since there is only one collective mode, $\nu=1$, $\gamma$ is a scalar and hence $\gamma^a=0$. Moreover, for energy-independent hybridization we have that $\partial_\epsilon G_R=-G_R^2$, and the A-matrix \eqref{Amatrix} can be written as~\bcite{BodePRL11}
\beq
A_\nu=-\pi WG_R[G_R,\Lambda_\nu]G_RW^\dagger\,.
\eeq 
Being the commutator of scalars, in this case $A_1=0$ and from Eq. \eqref{eq:damp2}, $\gamma^s$ must be positive and is given by Eq. \eqref{gseq}. (For an alternative derivation of the positiveness of the friction coefficient in a resonant-level system, see Ref. \bcite{HyldgaardMSEC03}). After some algebra, we obtain 
\beq
\left(\frac{\partial S}{\partial X}\right)^\dagger\frac{\partial S}{\partial X}=4\left[\frac{\partial \tilde \epsilon}{\partial X}\right]^2 \frac{\Gamma}{\left|{\mathcal L}\right|^2}
\left
( \begin{array}{cc} \Gamma_L & \sqrt{\Gamma_L\Gamma_R} 
 \\ \sqrt{\Gamma_L\Gamma_R} & \Gamma_R \end{array} \right)\,.
\eeq
and hence the damping coefficient becomes
\beq
\gamma(X)=-\int\frac{d\epsilon}{\pi}\Gamma\frac{\Gamma_L\partial_\epsilon f_L+\Gamma_R\partial_\epsilon f_R}{\left|{\mathcal L}\right|^4}\left[\frac{\partial \tilde \epsilon}{\partial X}\right]^2\,.
\eeq

\begin{figure}
\begin{center}
 \includegraphics[width=10cm]{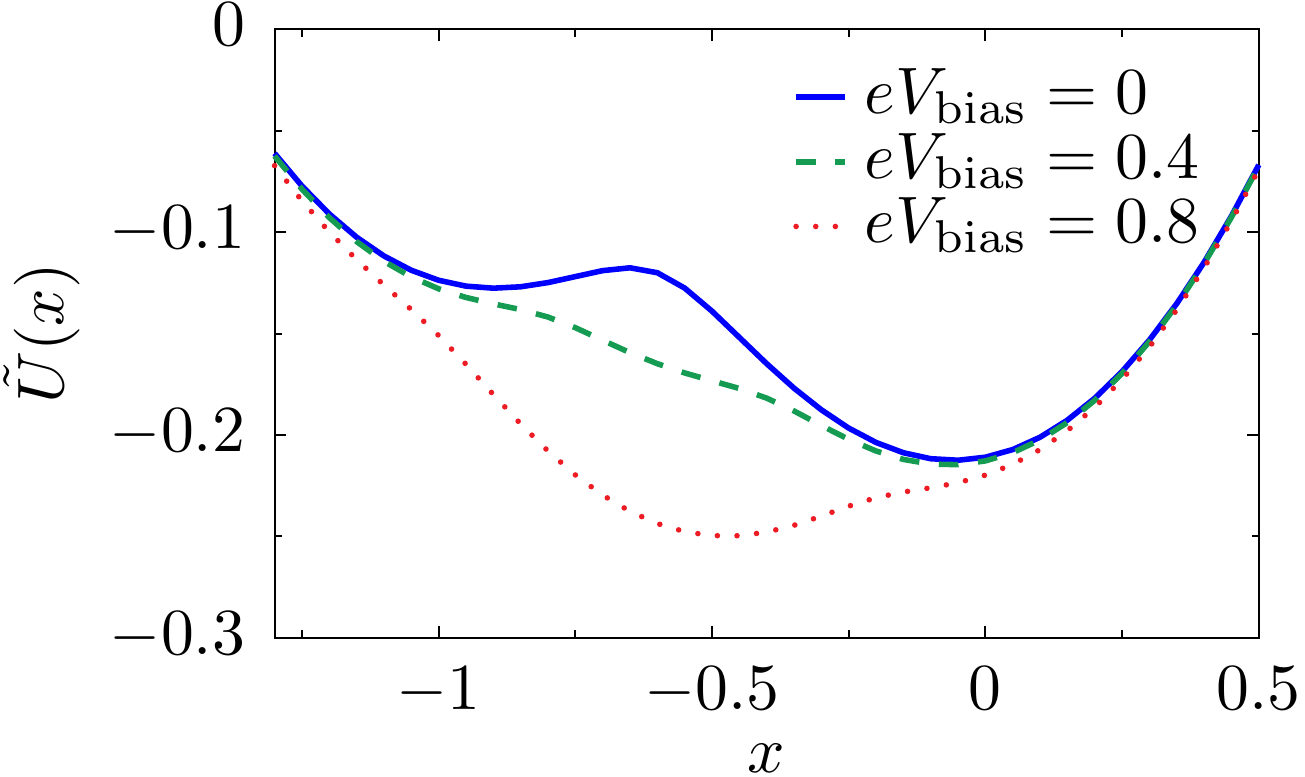}
\end{center}
\caption{Resonant level. The shape of the effective
potential $\tilde{U}(X)$ can be
tuned by the bias voltage. We consider the parameters $eV_{\mathrm{gate}}=0$, $\hbar \omega_0 = 0.01$ and
$\Gamma=0.1$. The dimensionless coordinate is $x = ({M
\omega_0^2}/{\lambda})
X$ and energies are measured in units of
${\lambda^2}/{(M\omega_0^2)}$.}
\label{fig:potential}
\end{figure}

We can evaluate the remaining integrals analytically in the zero-temperature limit \bcite{MozyrskyPRB06,PistolesiPRB08}. In the following we assume $\mu_L\ge\mu_R$. The average force is given by
\beq
  F(X) =-\frac{1}{\pi} \frac{\partial \tilde \epsilon}{\partial X}\sum_\alpha \frac{\Gamma_\alpha}{\Gamma}
\left[\arctan\left(\frac{\mu_\alpha -\tilde \epsilon}{\Gamma}\right)
+\frac{\pi}{2}\right]\,.
\eeq
Similarly we obtain the dissipation coefficient
\beq
  \gamma^s(X) = \frac{\Gamma}{\pi} \left[\frac{\partial \tilde \epsilon}{\partial X}\right]^2\sum_\alpha
\frac{\Gamma_\alpha}{\left[\left(\mu_\alpha-\tilde \epsilon\right)^2+\Gamma^2\right]^2}\,,
\eeq
together with the fluctuation
kernel 
\beq
  D(X)= \frac{ \Gamma_L \Gamma_R}{\pi \Gamma^3}\left[\frac{\partial \tilde \epsilon}{\partial X}\right]^2\left. \left[
\arctan\left(\frac{\mu-\tilde \epsilon}{\Gamma}\right) +
\frac{\Gamma(\mu-\tilde \epsilon)}{\left(\mu-\tilde \epsilon\right)^2+\Gamma^2}\right]\right|^{\mu=\mu_L}_{\mu=\mu_R}
\eeq
The position of the dot electronic level can be adjusted by an external gate voltage
\beq
eV_\mathrm{gate}=\frac{\mu_L+\mu_R}{2} - \epsilon_0\,,
\eeq
where the factor $(\mu_L+\mu_R)/2$ is included for convenience, to measure energies from the center of the conduction window. The difference in chemical potential between the leads is adjusted {\it via} a bias voltage
\beq
eV_\mathrm{bias}=\mu_L-\mu_R\,.
\eeq
For a single vibrational mode, the average current-induced force is necessarily conservative and we can define a corresponding potential. Restricting now our results to linear coupling, we write the local level as $\tilde \epsilon(X)=\epsilon_0+\lambda X$.  In Fig. \ref{fig:potential}, we show the effective potential $\tilde{U}(X) = \frac{M}{2} \omega_0^2 X^2 - \int dX F(X)$ which describes both the elastic and the current-induced forces at zero temperature and various bias voltages. Already this simple example shows that the current-induced forces can affect the mechanical motion qualitatively \bcite{PistolesiPRB08}. Indeed, the effective potential $\tilde U(X)$ can become multistable even for a purely harmonic elastic force and depends sensitively on the applied bias voltage.  

Alternative expressions of the current-induced forces for the resonant level model, in terms of phase shifts and transmission coefficients, are given in the Supplementary Material \ref{app23}.

\subsection{Two-level model}
\begin{figure}
\begin{center}
 \includegraphics[width=7.5cm]{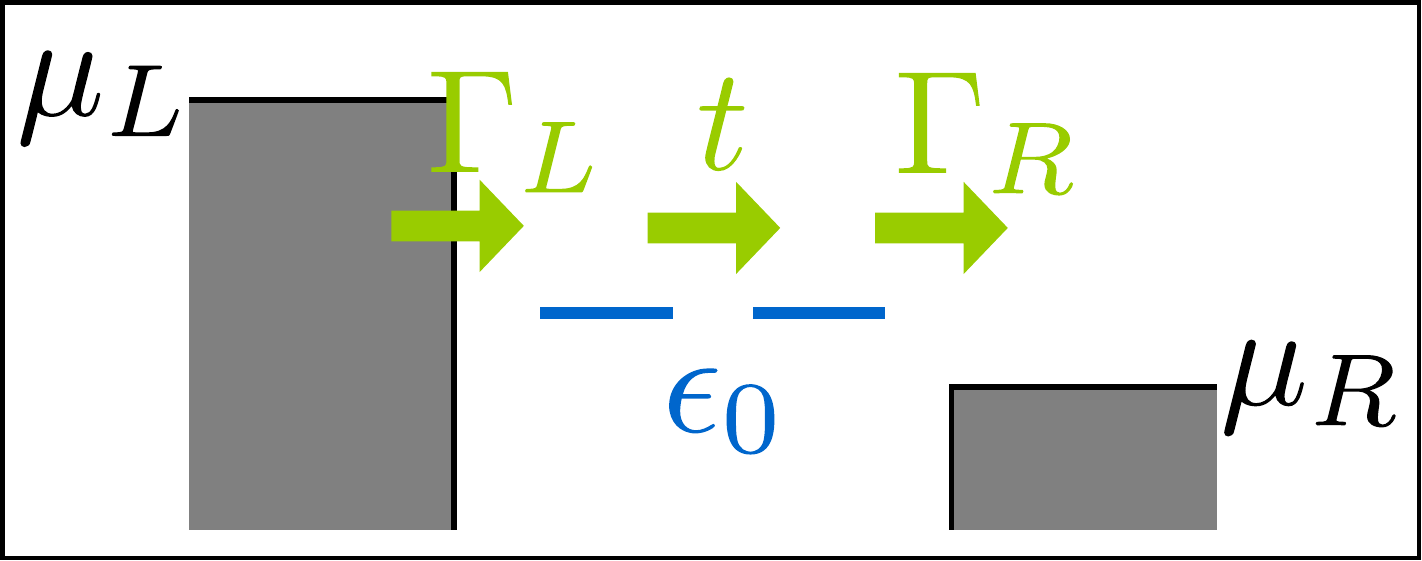}
\end{center}
\caption{Sketch of the two-level model. Electrons tunnel through two degenerate energy levels between left and right leads. The system is modulated by the coupling to the vibrational modes.}
\label{fig:sketch}
\end{figure}

For the resonant level model discussed so far, the A-matrix vanishes and the damping is necessarily positive. We now consider a model which allows for negative damping \bcite{Metelmann11}. Our toy model could be inspired by a double dot on a suspended carbon nanotube, or an H$_2$ molecule in a break junction. The model is depicted schematically in Fig. \ref{fig:sketch}.  The bare dot Hamiltonian
corresponds to degenerate electronic states $\epsilon_{0}$, localized on the left and right atoms or quantum dots, with tunnel coupling $t$ in between,
\begin{equation}\label{ToyHam}
H_0= 
\left( \begin{array}{cc} \epsilon_0&t\\
t & \epsilon_0\end{array}\right)\,.
\end{equation}
We consider a single
oscillator mode with coordinate $X$ that couples linearly to the difference in the occupation of the levels. In our previous notation, this means $\Lambda_1 = \lambda_1 \sigma_3$, where we denote by $\sigma_\mu$, with $\mu=0,\ldots,3$, the Pauli matrices acting in the two-site basis. The shift of the electronic levels is given by $\tilde\epsilon_{\pm}(X)=\epsilon_0\pm \lambda_1 X$.
The hybridization
matrices are given by
$\Gamma^{\alpha}=\frac{1}{2}\Gamma_{\alpha}(\sigma^{0}\pm\sigma^{3})$,
where the $+(-)$ refers to $\alpha=L(R)$. We can deduce the tunneling
matrix $W$ in terms of the hybridization matrices, 
\beq 
W=1/2 \sqrt{\Gamma_L/\pi}(\sigma^{0}+\sigma^{3})+1/2
\sqrt{\Gamma_R/\pi}(\sigma^{0}-\sigma^{3})\,.
\eeq
In the wide-band limit, we approximate $W$ and
$\Gamma_\alpha$ to be independent of energy. 
The retarded adiabatic GF takes the form
\begin{equation}
G^R(\epsilon,X) = \frac{1}{\Delta} 
\left( \begin{array}{cc} \epsilon-\tilde\epsilon_++i\Gamma_R &t\\
t & \epsilon-\tilde\epsilon_-+i\Gamma_L \end{array}\right)\,,
\end{equation}
with $\Delta(X)=\left(\epsilon-\tilde\epsilon_-+ i\Gamma_L)(\epsilon-\tilde\epsilon_++i\Gamma_R\right)-t^2$.

For simplicity, we restrict our attention to symmetric couplings to the leads, $\Gamma_L=\Gamma_R=\Gamma/2$.
Hence the frozen S-matrix $S(\epsilon,X)$ becomes
\begin{equation}
S(\epsilon,X) =1- \frac{i \Gamma}{\Delta} 
\left( \begin{array}{cc}
\epsilon-\tilde\epsilon_++i\Gamma/2
&t \Gamma\\
t \Gamma& \epsilon-
\tilde\epsilon_-  +i\Gamma/2 \end{array}\right)\,,
\end{equation}
while the A-matrix takes the form
\beq
A(\epsilon,X)=i\lambda_1 \Gamma\,t\,\frac{\left(\epsilon-\epsilon_0+i\Gamma/2\right)^2 +
i\left[(\lambda_1 X)^2-t^2\right]}{\Delta^3}\,\sigma_2\,.
\eeq

We can now give explicit expressions for the current-induced forces. The explicit expressions are lengthy and are given in Supp. Mat. \ref{app23b}, Eqs. \eqref{MeanF2L} and \eqref{Gamma2L} for the mean force and damping matrix, respectively. The variance of the fluctuating force can be calculated accordingly.

\begin{figure}
\begin{center} 
\includegraphics[height=4.5 cm,keepaspectratio=true]{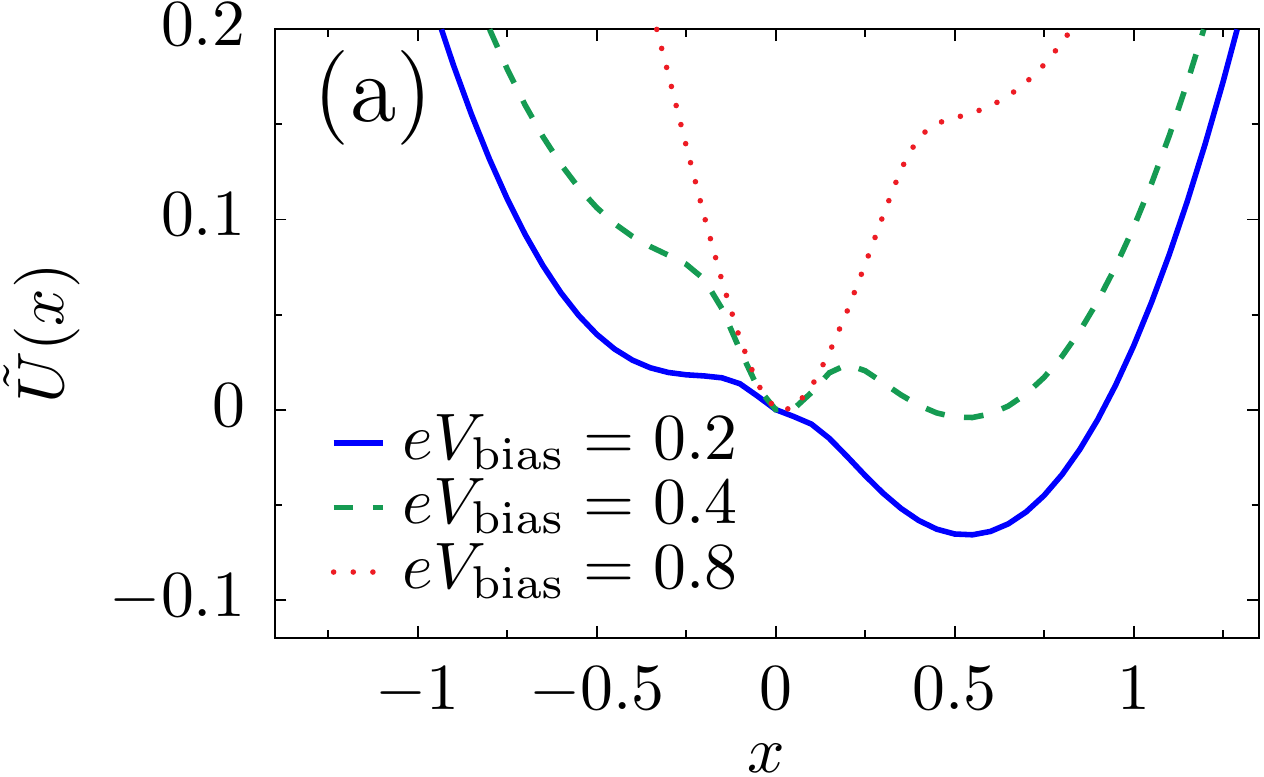}
\includegraphics[height=4.5 cm,keepaspectratio=true]{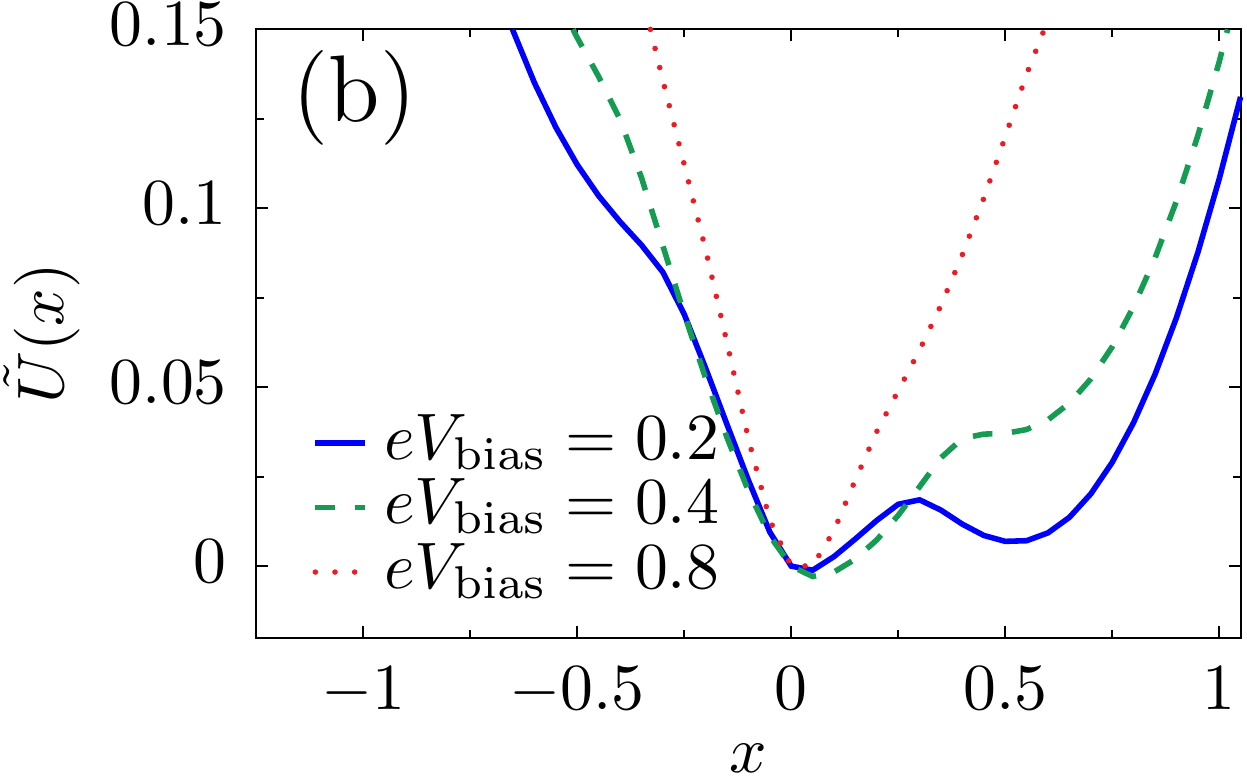}
\includegraphics[height=4.5 cm,keepaspectratio=true]{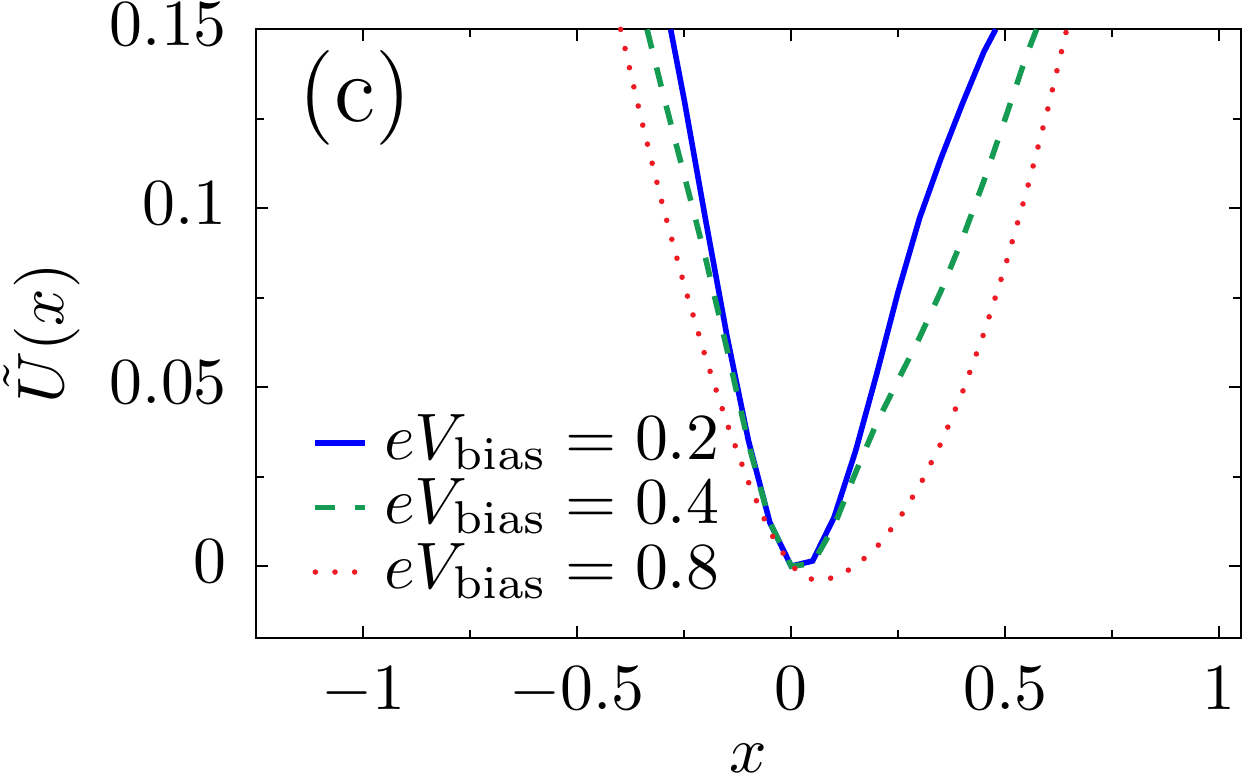}
\end{center}
\caption{Effective potential for the mechanical motion in the two-level model. The shape of the potential can be tuned by changing the
bias and gate voltages: (a) $eV_{\mathrm{gate}}=0$,
(b) $eV_{\mathrm{gate}}=0.2$ and (c) $eV_{\mathrm{gate}}=0.4$. We consider the parameters $\hbar \omega_0 = 0.01$,
$t=0.1$ and
$\Gamma=0.1$. The dimensionless coordinate is $x = ({M
\omega_0^2}/{\lambda_1})
X$ and energies are measured in units of
${\lambda_1^2}/{(M\omega_0^2)}$.}
\label{fig:potential2}
\end{figure}

The average force given in Eq. \eqref{MeanF2L} of Supp. Mat. \ref{app23b} combines with the elastic force to give rise to the effective potential $\tilde U(X)$ depicted, for zero temperature, in Fig. \ref{fig:potential2}. As in the case studied in the previous section, the system can exhibit various levels of multistability when changing the bias.

\begin{figure}
\begin{center} 
\includegraphics[height=4.5 cm,keepaspectratio=true]{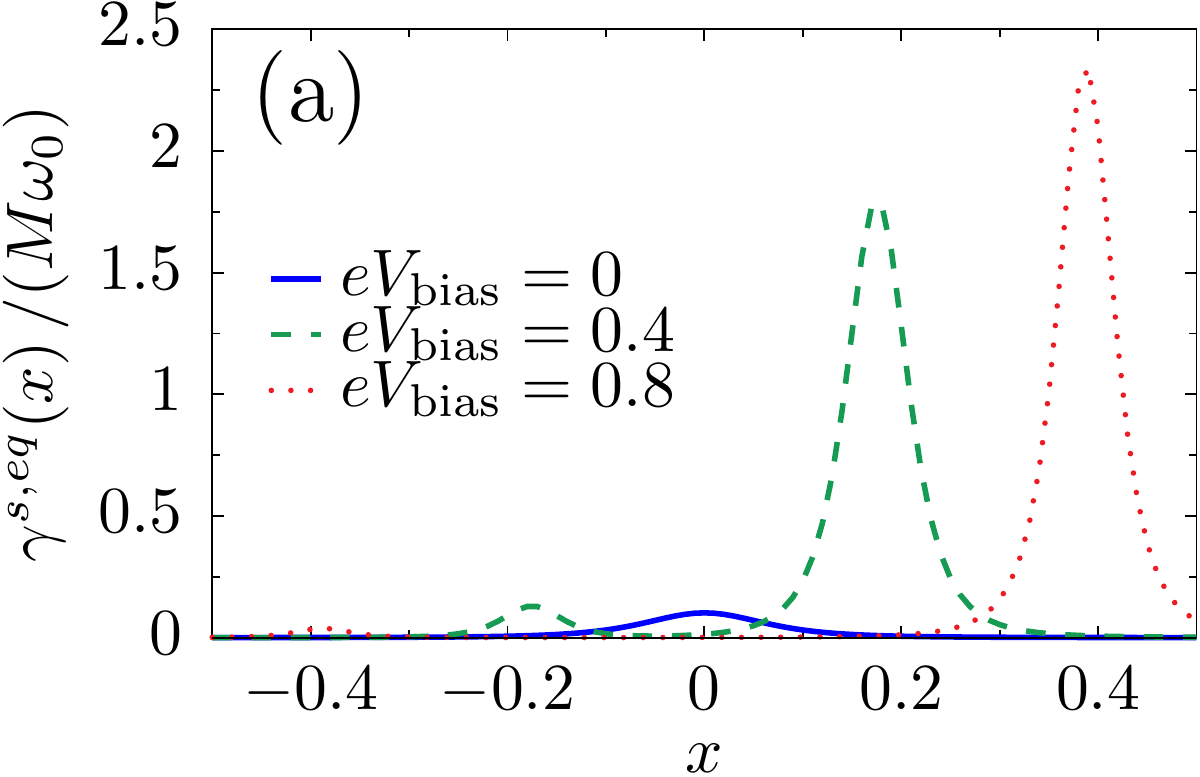}
\includegraphics[height=4.5 cm,keepaspectratio=true]{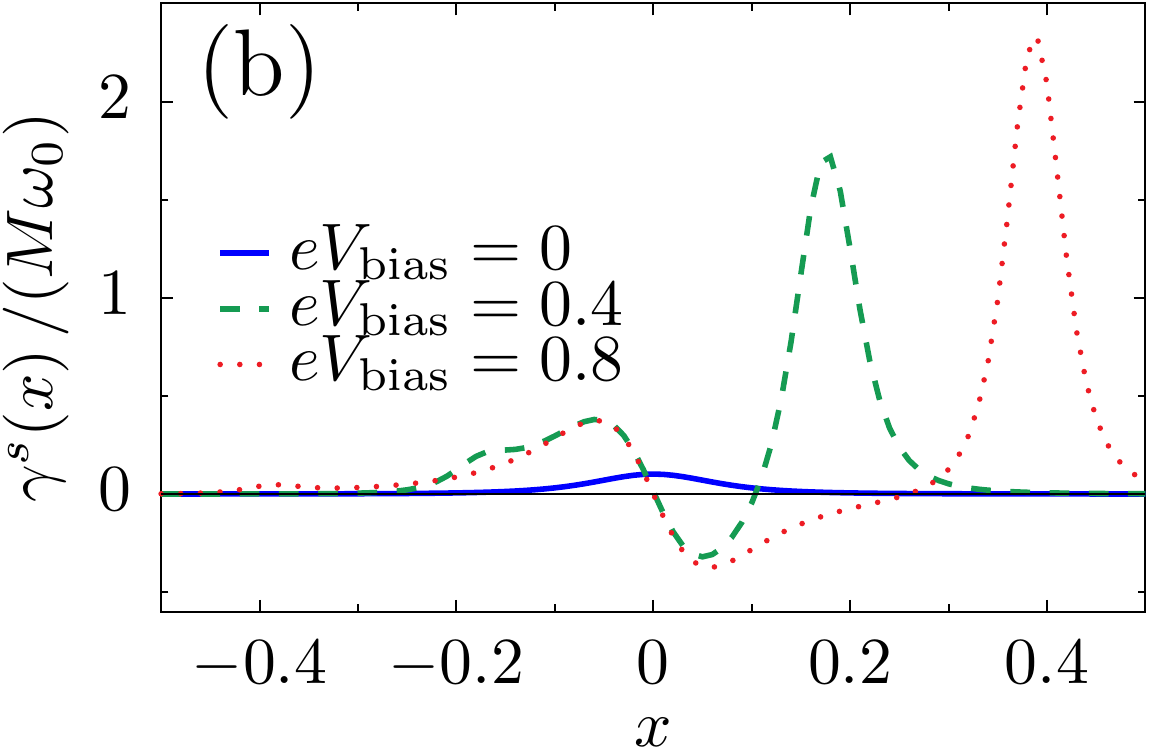}
\includegraphics[height=4.5 cm,keepaspectratio=true]{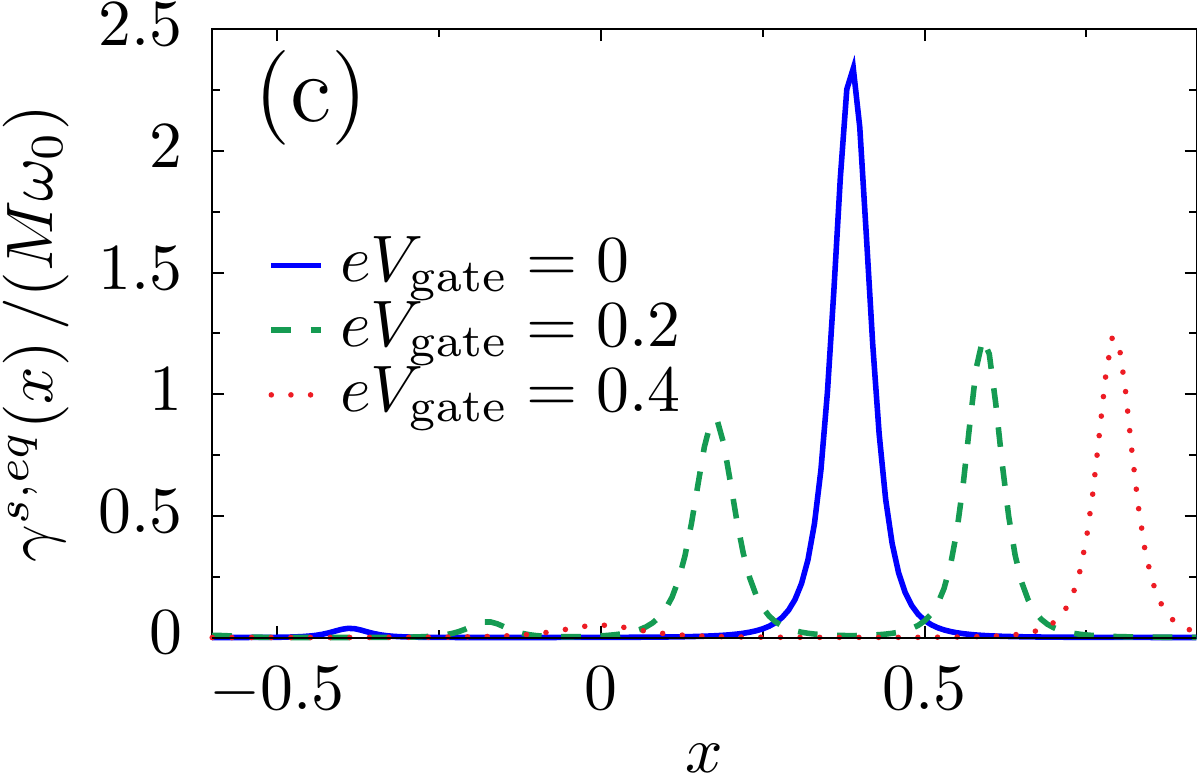}
\includegraphics[height=4.5 cm,keepaspectratio=true]{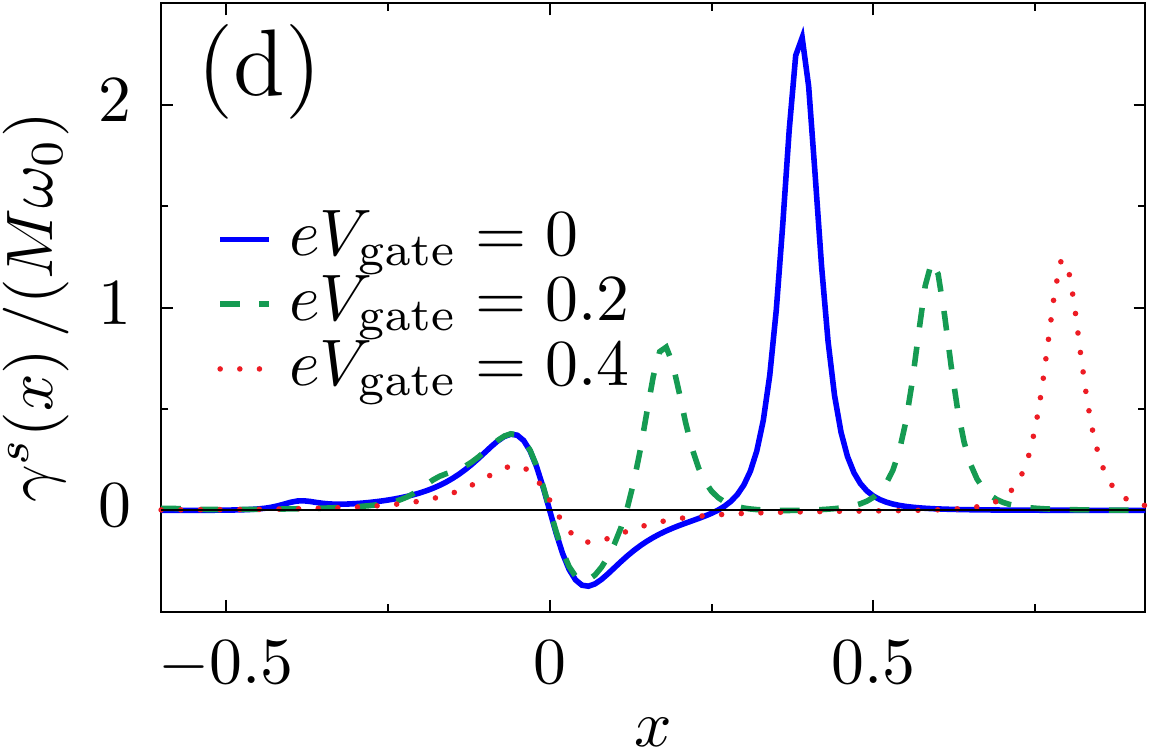}
\end{center}
\caption{Damping {\it vs.} mechanical displacement in the two-level model. (a) Contribution $\gamma^{s,eq}$ to the friction coefficient for various bias voltages at fixed gate
voltage $eV_{\mathrm{gate}}=0$. (b) At the same gate voltage, the total damping exhibits a
region of negative damping due to the contribution of $\gamma^{s,ne}$.
(c) $\gamma^{s,eq}$ for various gate
voltages with the bias voltage $eV_{\mathrm{bias}}=0.8$.  Note that for both $eV_{\mathrm{gate}}=0.2$ and $eV_{\mathrm{gate}}=0.4$, one small peak for negative $x$ falls outside of the shown range of $x$. (d) Again, the full damping $\gamma^{s}$ exhibits regions of negative damping. We choose $\hbar \omega_0 = 0.01$, $\Gamma=0.1$ and $t=0.1$. The dimensionless coordinate is $x = ({M \omega_0^2}/{\lambda_1}) X$ and energies are measured in units of ${\lambda_1^2}/{(M\omega_0^2)}$.}
\label{fig:tm2damping1}
\end{figure}

\begin{figure}
\begin{center} 
\includegraphics[height=3 cm,keepaspectratio=true]{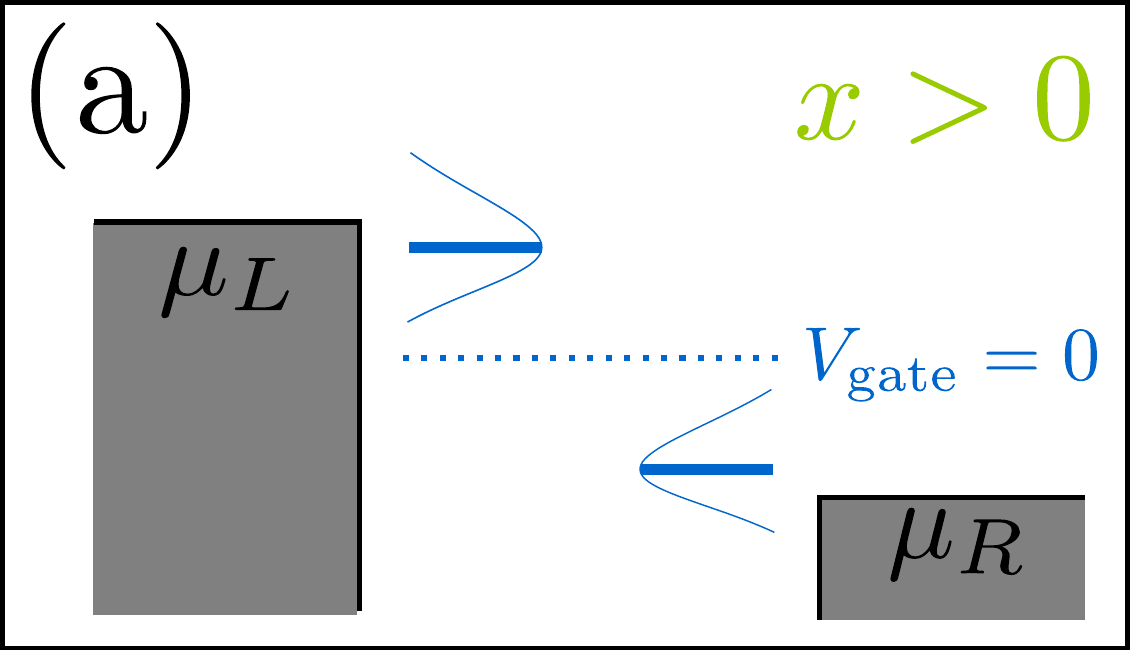}
\includegraphics[height=3 cm,keepaspectratio=true]{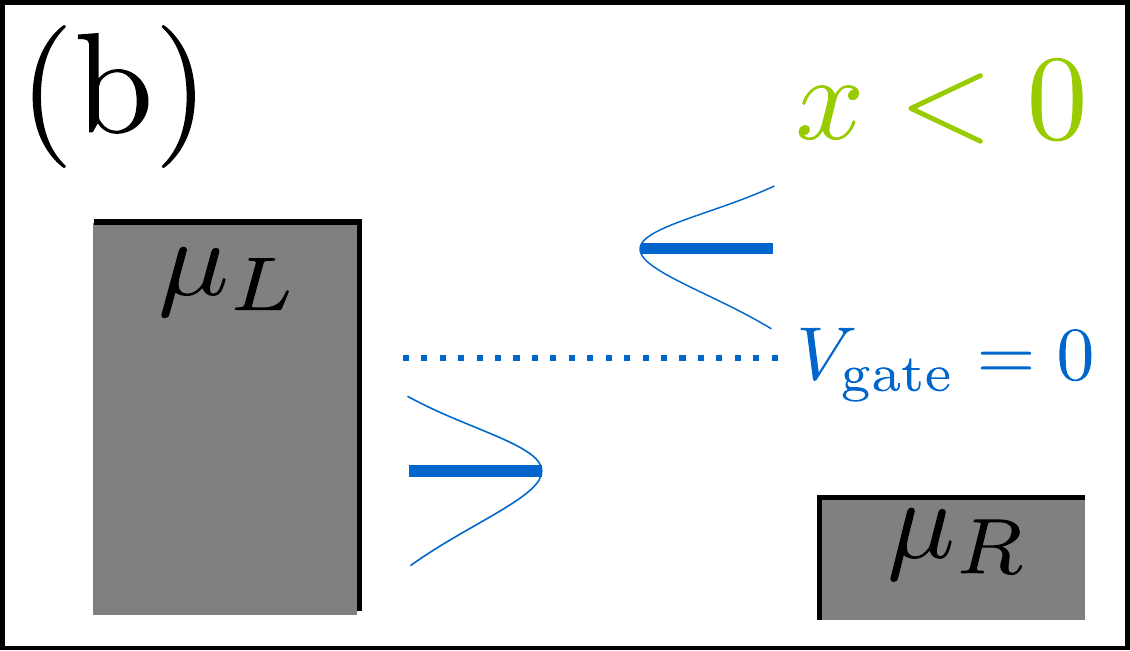}
\includegraphics[height=3 cm,keepaspectratio=true]{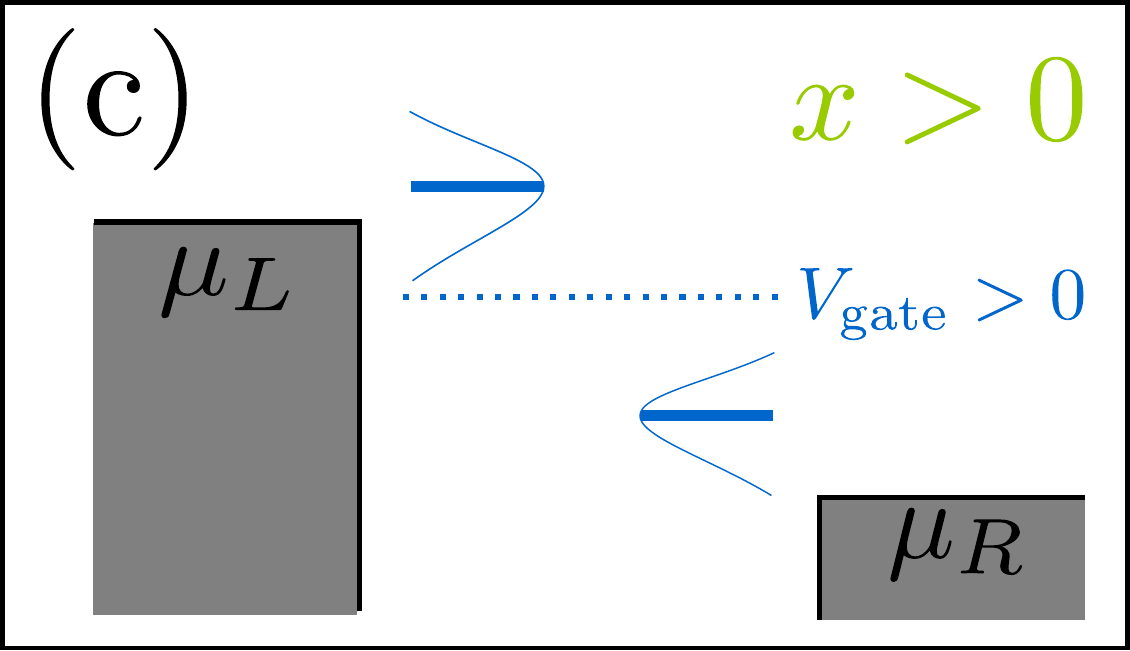}
\includegraphics[height=3 cm,keepaspectratio=true]{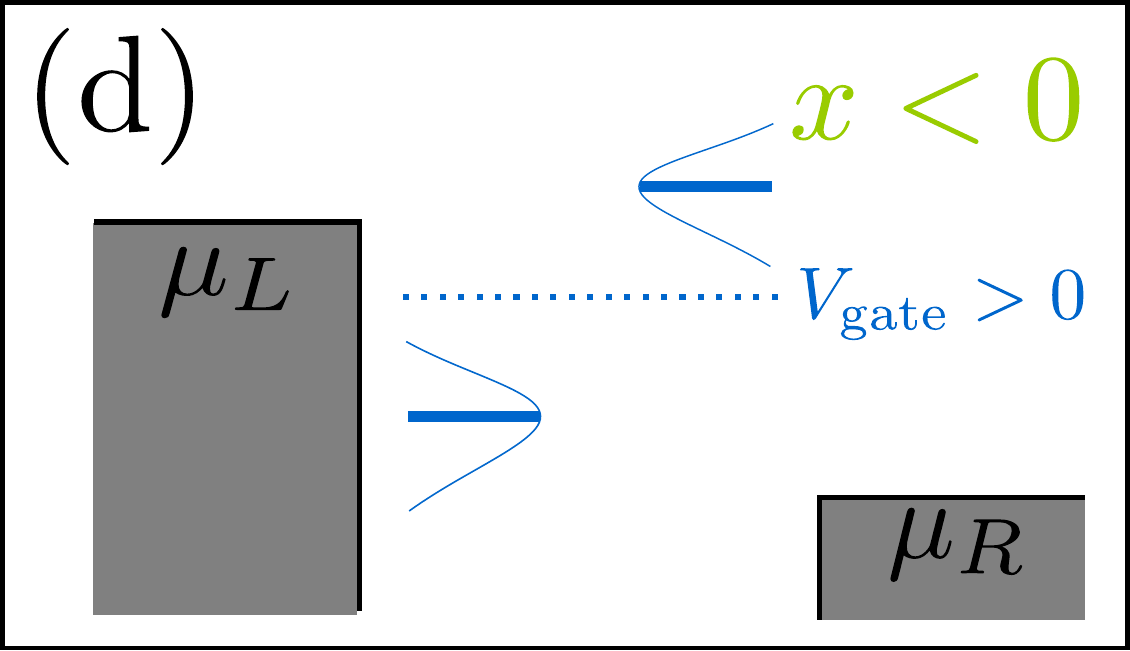}
\end{center}
\caption{Cartoon of the positions of the electronic levels in the dot with respect to the Fermi levels of the leads, depending on the sign of $x$ and the existence of a gate voltage. The levels are broadened due to the hybridization with the leads $\Gamma$. When $x>0$, ``left'' and ``right'' levels approach the Fermi levels of left and right leads respectively, (a) for $eV_{\mathrm{gate}}=0$ the levels align simultaneously for left and right, (c) a finite $eV_{\mathrm{gate}}$ produces an assymmetry between left and right. For $x<0$ the alignment of the levels is inverted, (b) $eV_{\mathrm{gate}}=0$, (d) finite $eV_{\mathrm{gate}}$.}
\label{fig:nielsfig}
\end{figure}

The results for the friction coefficient, given in Supp. Mat. \ref{app23b} Eq. \eqref{Gamma2L}, are shown in Fig. \ref{fig:tm2damping1} as a function of the dimensionless oscillator coordinate $x$, for zero temperature. The contribution $\gamma^{s,eq}$ to the friction coefficient is peaked at $e V_{\mathrm{gate}}\pm eV_{\mathrm{bias}}/2= \pm\sqrt{(\lambda_1 X)^2 +t^2}$, as depicted in Figs. \ref{fig:tm2damping1} (a) and (c). Neglecting the coupling to the leads, our toy model can be considered as a two-level system with level-spacing $2\sqrt{(\lambda_1 X)^2 + t^2}$. Thus, the peaks occur when one of the dot's electronic levels enters the conduction window. When this happens, small changes in the oscillator coordinate $X$ can have a large impact on the occupation of the levels. This effect is more pronounced when the dots' levels pass the Fermi levels that they are directly attached to [corresponding to $X>0$ for current flowing from left to right, see Fig. \ref{fig:tm2damping1} (a) and Fig.  \ref{fig:nielsfig} (a), (b)]. The broadening of the peaks is due to the hybridization with the leads, $\Gamma/2$. When $e V_{\mathrm{gate}}=0$, two peaks are expected symmetrically about $X=0$, as shown in  Fig. \ref{fig:tm2damping1} (a) [see also Figs.  \ref{fig:nielsfig} (a) and (b)]. The effect of a finite gate voltage $e V_{\mathrm{gate}}$ is two-fold: it shifts the non-interacting electronic levels of the dot away from the middle of the conduction window, and hence the shifted levels $\tilde\epsilon_\pm$ pass the Fermi levels of right and left leads at different values of $X$, Figs.  \ref{fig:nielsfig} (c) and (d). Therefore in this case four peaks are expected, with two larger peaks located at $X>0$, and two smaller peaks located at $X<0$. This is shown in Fig. \ref{fig:tm2damping1} (c). The height of the peaks in this case is reduced with respect to the case $e V_{\mathrm{gate}}=0$, since for a given peak, only one of the dot's levels is in resonance with one of the leads. Note that four real values of $X$ can be obtained only if $\left(e V_{\mathrm{gate}}\pm eV_{\mathrm{bias}}/2\right)^2>t^2$. A situation with $\left(e V_{\mathrm{gate}}- eV_{\mathrm{bias}}/2\right)^2<t^2$ while $\left(e V_{\mathrm{gate}}+ eV_{\mathrm{bias}}/2\right)^2>t^2$ is shown in \ref{fig:tm2damping1} (c) (red-dotted line), where a big peak is observed for $X=1/\lambda_1\sqrt{\left(e V_{\mathrm{gate}}+ eV_{\mathrm{bias}}/2\right)^2-t^2}$, a corresponding small peak for $X=-1/\lambda_1\sqrt{\left(e V_{\mathrm{gate}}+ eV_{\mathrm{bias}}/2\right)^2-t^2}$ [not displayed in Fig. \ref{fig:tm2damping1} (c)], plus a peak at $X=0$.

\begin{figure}
\begin{center} 
\includegraphics[height=4.5 cm,keepaspectratio=true]{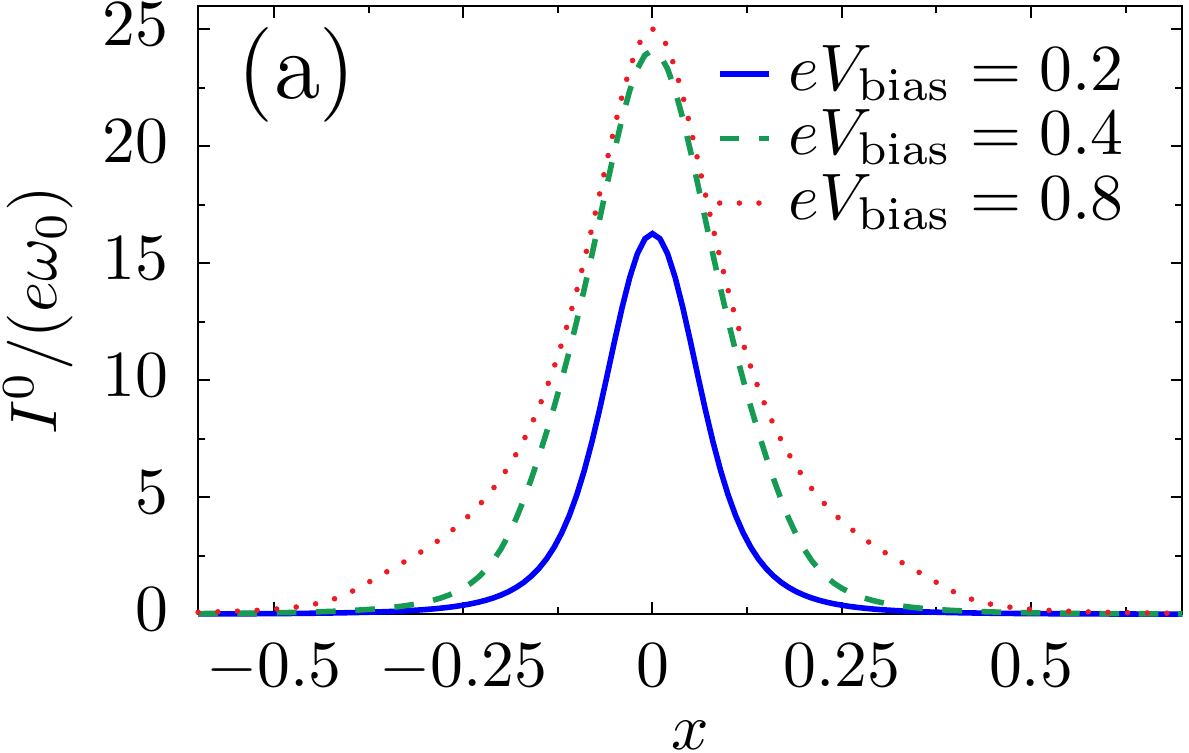}
\includegraphics[height=4.5 cm,keepaspectratio=true]{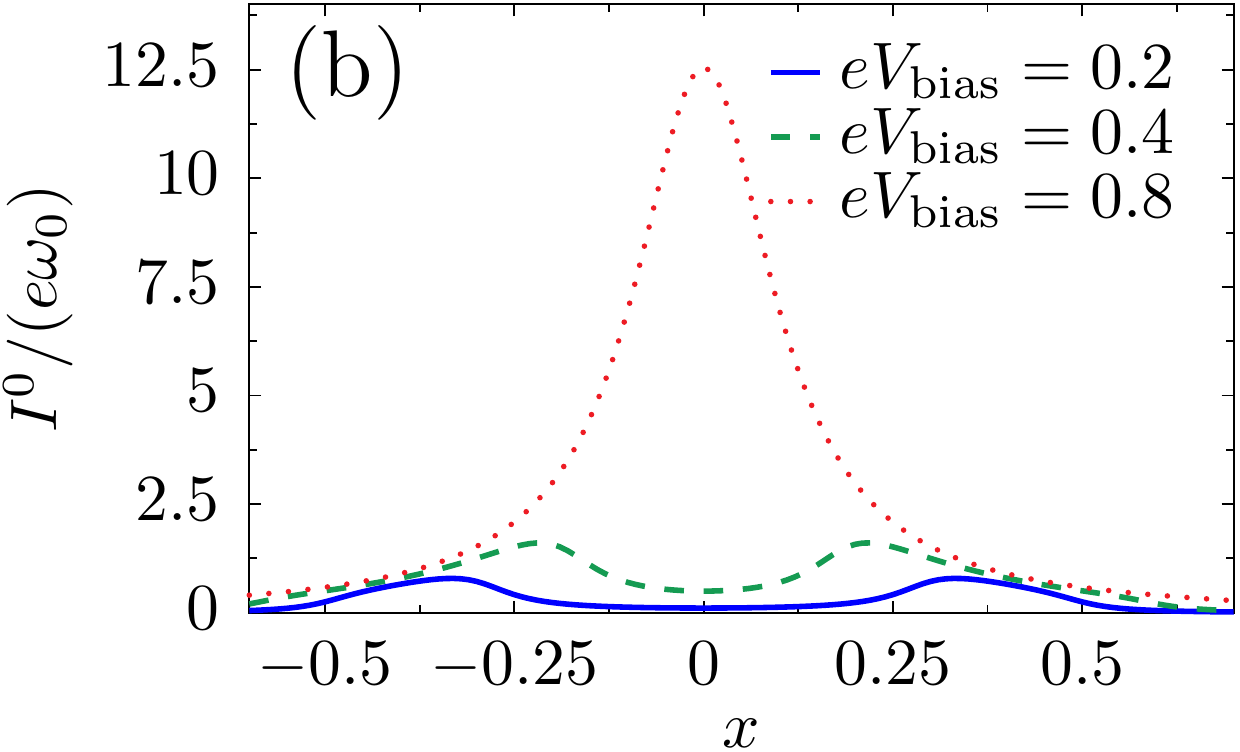}
\includegraphics[height=4.5 cm,keepaspectratio=true]{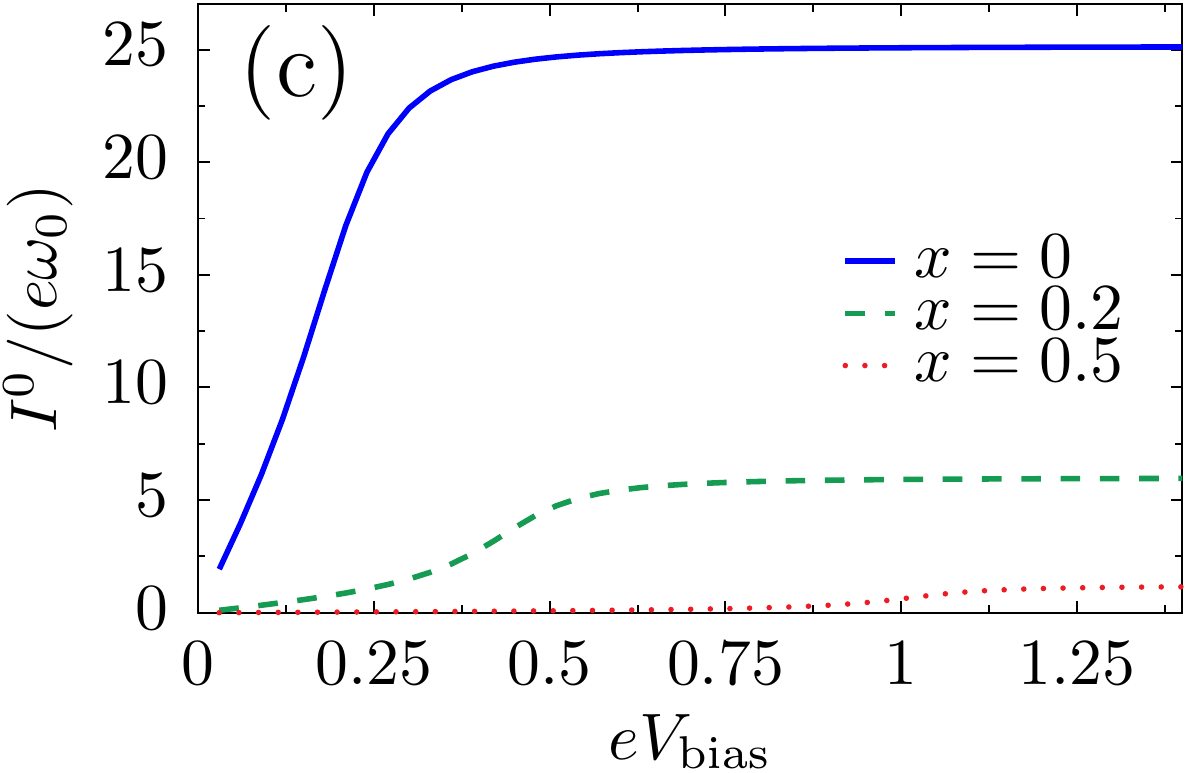}
\includegraphics[height=4.5 cm,keepaspectratio=true]{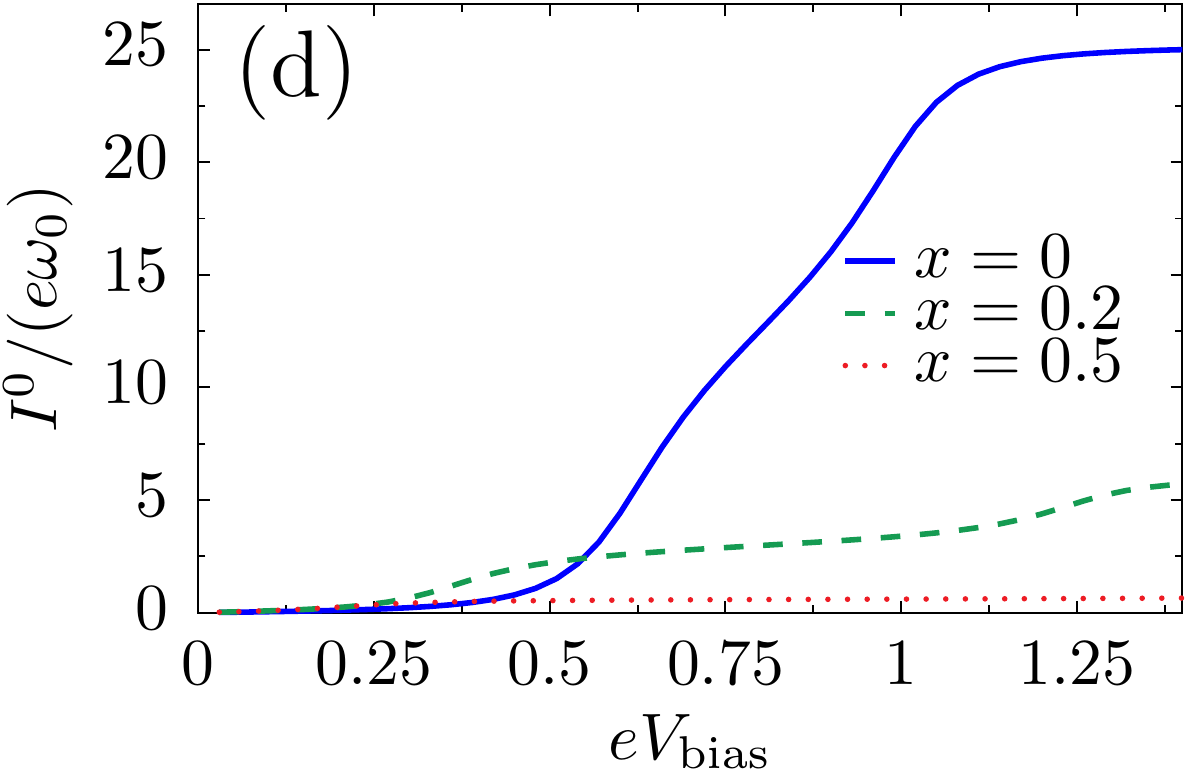}
\includegraphics[height=4.5 cm,keepaspectratio=true]{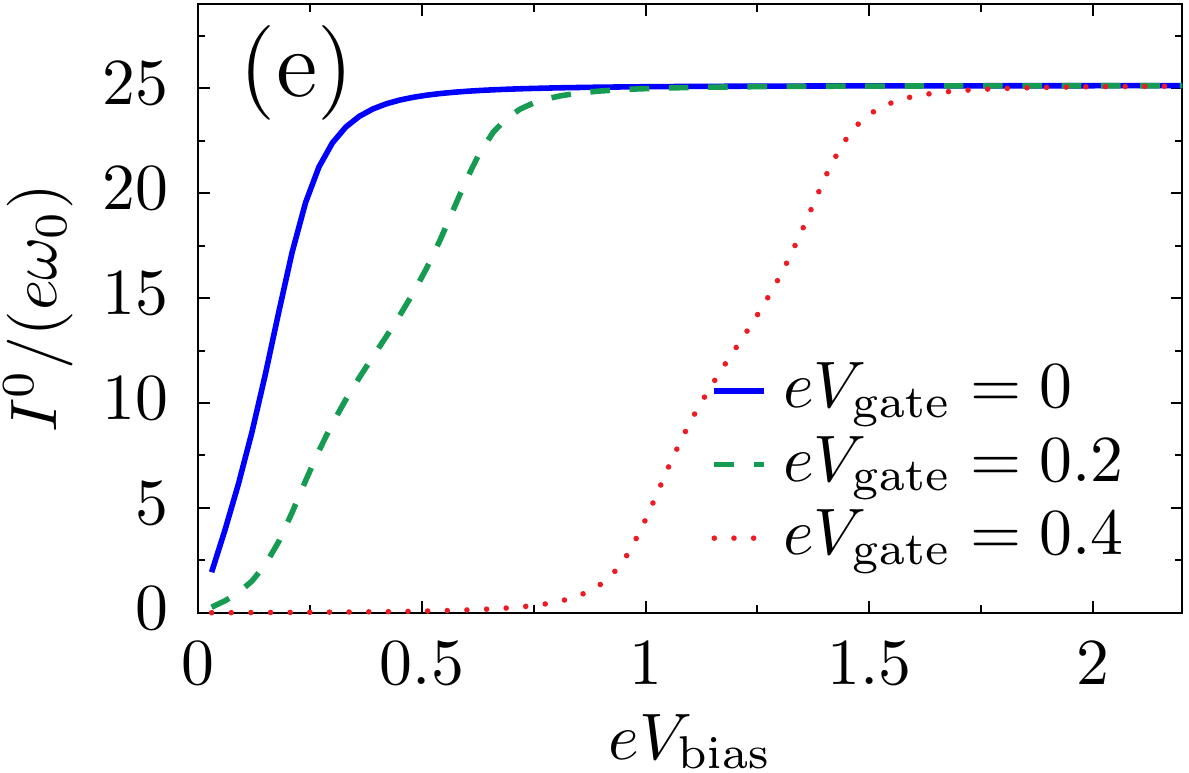}
\includegraphics[height=4.5 cm,keepaspectratio=true]{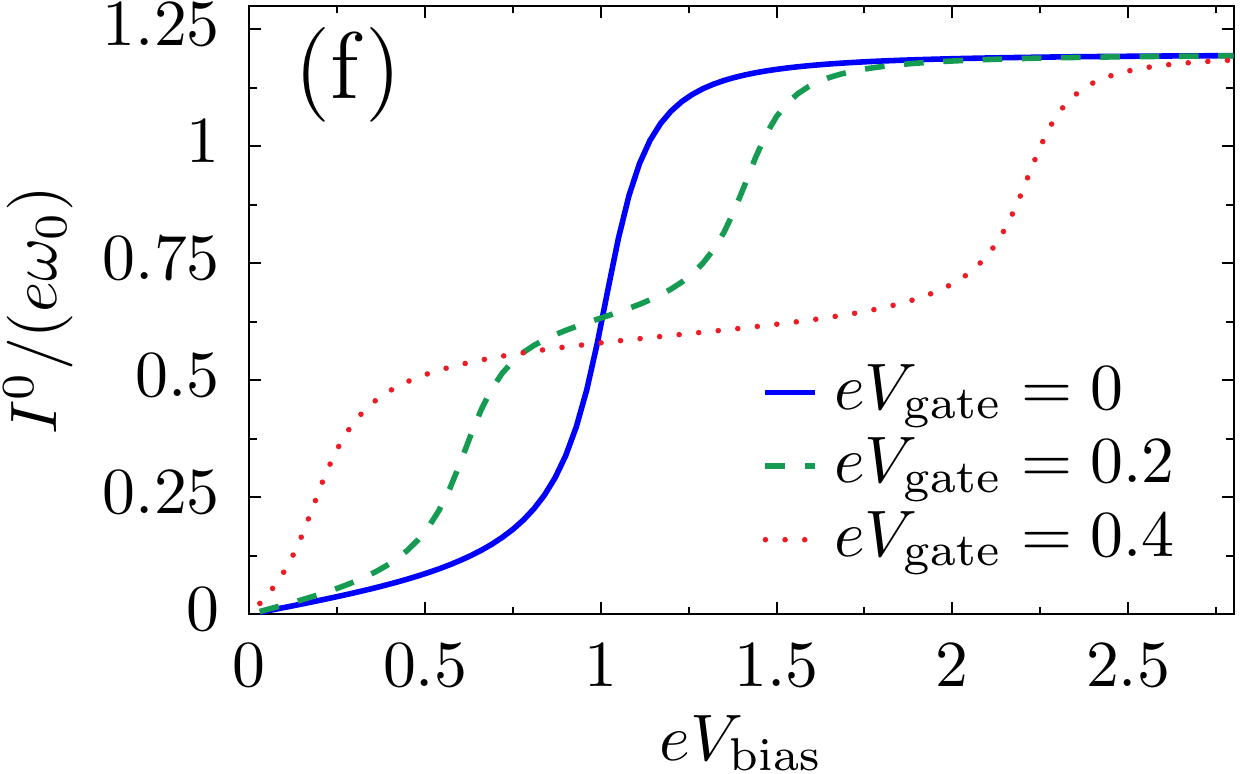}
\end{center}
\caption{Dependence of the current in the two-level model on various parameters. Current as function of mechanical displacement for (a) $V_\mathrm{gate}=0$ and (b) $V_\mathrm{gate}=0.4$; as function of bias for (c) $V_\mathrm{gate}=0$, (d) $V_\mathrm{gate}=0.4$, (e) $x=0$ and (f) $x=0.5$. We choose $\hbar \omega_0 = 0.01$, $\Gamma=0.1$ and $t=0.1$. The dimensionless coordinate is $x = ({M
\omega_0^2}/{\lambda_1})
X$ and energies are measured in units of
${\lambda_1^2}/{(M\omega_0^2)}$.}
\label{fig:tm2current}
\end{figure}

For this model, the A-matrix is generally non-vanishing, which can result in negative damping for out-of-equilibrium situations. This is due to a negative contribution of $\gamma^{s,ne}$ to the total damping. This is visualized in Figs.
\ref{fig:tm2damping1} (b) and (d). Negative damping is possible when both dot levels are inside the conduction window, restricting the region in $X$ over which negative damping can occur. Indeed, when only one level is within the conduction window, the system effectively reduces to the resonant level model for which, as we showed in the previous subsection, the friction coefficient $\gamma^{s}$ is always positive.  When current flows from left to right, negative damping occurs only for positive values of the oscillator coordinate $X$, as shown in Figs. \ref{fig:tm2damping1} (b) and (d). This is consistent with a level-inversion picture, as discussed recently in Ref. \bcite{LuePRL11}. Pictorially, the electron-vibron coupling causes a splitting in energy of the left and right levels.  When $X>0$, electrons can go ``down the ladder'' formed by the energy levels by passing energy to the oscillator and hence amplifying the vibrations. For $X<0$, electrons can pass between the two dots only by absorbing energy from the vibrations, causing additional non-equilibrium damping.  For small broadening of the dot levels due to the coupling to the leads, this effect is expected to be strongest when the vibration-induced splitting $\lambda_1 X$ becomes of the same order as the strength of the hopping $t$. When $X$ grows further, the increasing detuning of the dot levels reduces the current and hence the non-equilibrium damping [see Figs. \ref{fig:tm2damping1} (b) and (d) and Figs. \ref{fig:tm2current} (a), (b)].
The coexistence of a multistable potential together with regions of negative damping can lead to interesting nonlinear behavior for the dynamics of the oscillator. In particular, and as we show in the next example, limit-cycle solutions are possible, in the spirit of a Van der Pol oscillator \bcite{HanggiAJP83}.

We can also calculate the current. The pumping contribution is proportional to the velocity $\dot X$ and thus small. Therefore we show here results only for the dominant adiabatic part of the current. This is given by
\begin{align}
  I^0 = \frac{e}{h} \int d\epsilon \frac{2 t^2 \Gamma^2 (f_L-f_R)}{\left\vert
\Delta\right\vert^2}\,.
\end{align}
For zero temperature, the behavior of the current is shown in Fig. \ref{fig:tm2current} as a function of various parameters. Figs. \ref{fig:tm2current} (a) and (b) show the current as a function of the (dimensionless) oscillator coordinate $x$ for two different values of gate potential for which the system exhibits multistability by developing several metastable equilibrium positions. For $V_\mathrm{gate}=0$ and independently of bias, the current shows a maximum at the local minimum of the effective potential $x=0$, while $I^0\approx 0$ for another possible local minimum, $x\approx 0.5$ (compare with Fig. \ref{fig:potential2} (a)). The true equilibrium value of $x$ can be tuned {\it via} the bias potential, showing the possibility of perfect switching. For finite gate potential however, the current is depleted from $x=0$ with diminishing bias. Figs. \ref{fig:tm2current} (c) to (d) show the current as a function of gate or bias voltage for fixed representative values of the oscillator coordinate $x$.  The
current changes stepwise as the number of levels inside the conduction window changes, coinciding with the peaks in the friction coefficient illustrated in Fig. \ref{fig:tm2damping1}. In an experimental setting, the measured {\it dc} current would involve an average over the probability distribution of the coordinate $x$, given by the solution of the Fokker-Planck equation associated to the Langevin equation \eqref{langevin}.

\subsection{Two vibrational modes}
As a final example, we present a simple model which allows for both a non-conservative force and an effective ``Lorentz'' force, in addition to negative damping. For this it is necessary to couple the two electronic orbitals of the previous example, see Eq. \eqref{ToyHam}, to at least two oscillatory modes which we assume to be degenerate. 
The relevant vibrations in this case can be thought of as a center-of-mass
vibration $X_1$ between the leads, and a stretching mode $X_2$. (It should be noted that this is for visualization purposes only. In reality, for an H$_2$ molecule, the stretching mode is a high energy mode when compared to a transverse and a rotational mode, see Ref. \bcite{DjukicPRB05}. Nevertheless, the H$_2$ molecule does indeed have two near-degenerate low energy vibrational modes, corresponding to rigid vibrations between the leads and a rigid rotation relative to the axis defined by the two leads.) The stretch mode 
modulates the hopping parameter,
\beq\label{tH2}
t\to \tilde t(X_2) = t+\lambda_2X_2\,,
\eeq
while the center of mass mode $X_1$ is modeled as coupling linearly to the density,
\beq\label{epsH2}
\epsilon_0\to \tilde \epsilon(X_1)=\epsilon_0
+\lambda_1 X_1\,,
\eeq
hence $\Lambda_1=\lambda_1 \sigma_0$ and $\Lambda_2=\lambda_2\sigma_1$. We work in the wide-band limit, but allow for asymmetric coupling to the leads. The retarded Green's function becomes
\begin{equation}
G^R(\epsilon,X_1,X_2) = \frac{1}{\Delta} 
\left( \begin{array}{cc} \epsilon-\tilde \epsilon+i\Gamma_R &\tilde t\\
\tilde t & \epsilon-\tilde \epsilon+i\Gamma_L \end{array}\right)\,,
\end{equation}
where now $\Delta(X_1,X_2)=(\epsilon-\tilde\epsilon+i\Gamma_L)(\epsilon-\tilde\epsilon+i\Gamma_R)-\tilde t^2$.
The frozen S-matrix can be easily calculated to be
\begin{equation}
S(\epsilon,X_1,X_2) =1- \frac{2i}{\Delta} 
\left( \begin{array}{cc} \left(\epsilon-\tilde \epsilon+i\Gamma_R\right)\Gamma_L
&\tilde t \sqrt{\Gamma_L\Gamma_R}\\
\tilde t\sqrt{\Gamma_L\Gamma_R} & \left(\epsilon-\tilde
\epsilon+i\Gamma_L\right)\Gamma_R  \end{array}\right)\,.
\end{equation}
The A-matrices also take a
simple form for this model. Since $\Lambda_1$ is proportional to the identity
operator, 
\beq\label{A1H2}
A_1(\epsilon,X_1,X_2)=-\pi\lambda_1\,WG_R\,[G_R,\sigma_0]\,G_RW^\dagger=0\,.
\eeq
On the other hand, the A-matrix associated with $X_2$ is non-zero and given by 
\beq\label{A2H2}
A_2(\epsilon,X_1,X_2)=-i\lambda_2\,\frac{\sqrt{\Gamma_1\Gamma_2}}{\Delta^2}\,\sigma_2\,.
\eeq
From this we can compute the average force, damping, pseudo-Lorentz force, and noise terms. These are listed in Supp. Mat. \ref{app3}. At zero temperature, it is possible to obtain
analytical expressions for these current-induced forces. 
Studying the dynamics of the modes $X_{1,2}(t)$ implies solving the two coupled
Langevin equations given by Eq. \eqref{langevin}, after inserting the
expressions for the forces given in Supp. Mat. \ref{app3}. Within our formalism we are
able to study the full non-linear dynamics of the problem, which brings out a
plethora of new qualitative behavior. In particular, analyses which linearize the current-induced force about a static equilibrium point would predict run-away modes due to negative
damping and non-conservative forces \bcite{LueNanoLett10}. Taking into account nonlinearities allows one to find
the new stable attractor of the motion. Indeed, we find that these linear instabilities typically result in dynamical equilibrium, namely limit-cycle dynamics \bcite{BodePRL11}. We note in passing that limit cycle dynamics in a nanoelectromechanical system was also discussed recently in Ref. ~\bcite{Metelmann11}. 

\begin{figure}
\begin{center}
\includegraphics[width=6 cm,keepaspectratio=true]{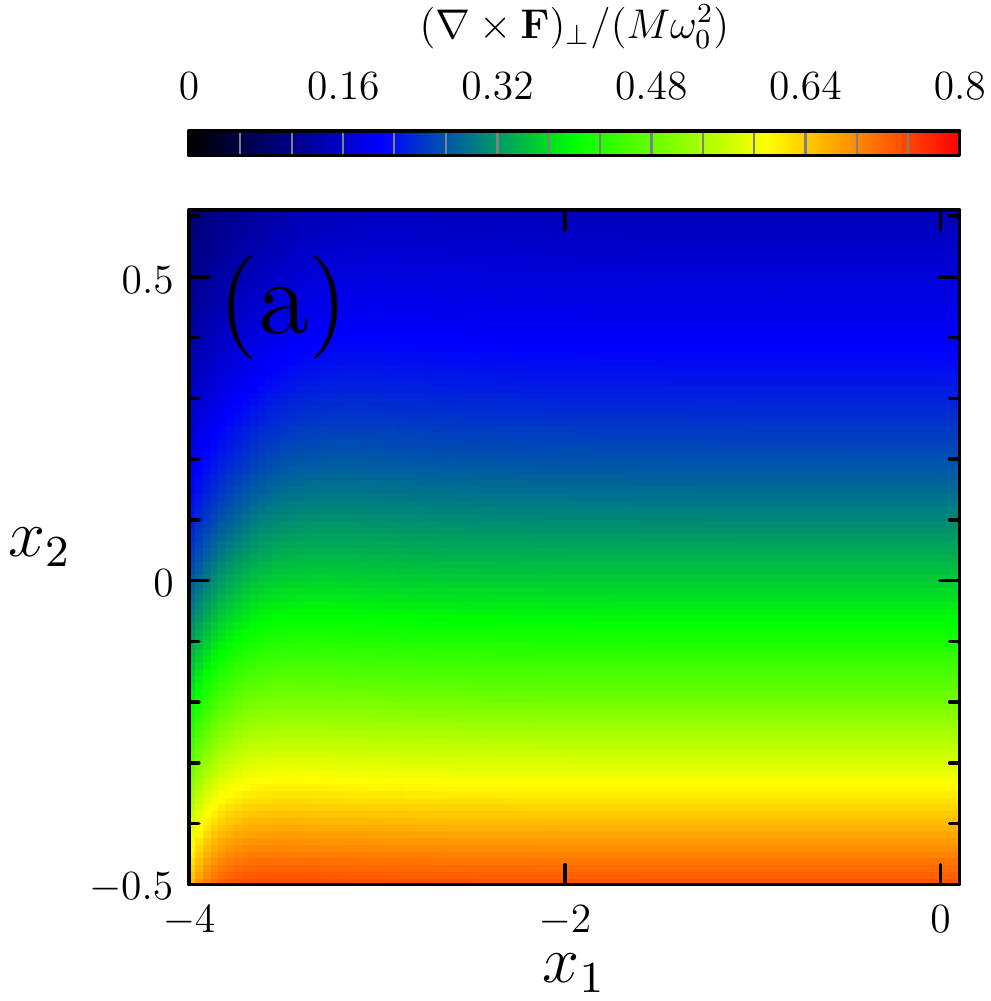}\, \,
\includegraphics[width=6 cm,keepaspectratio=true]{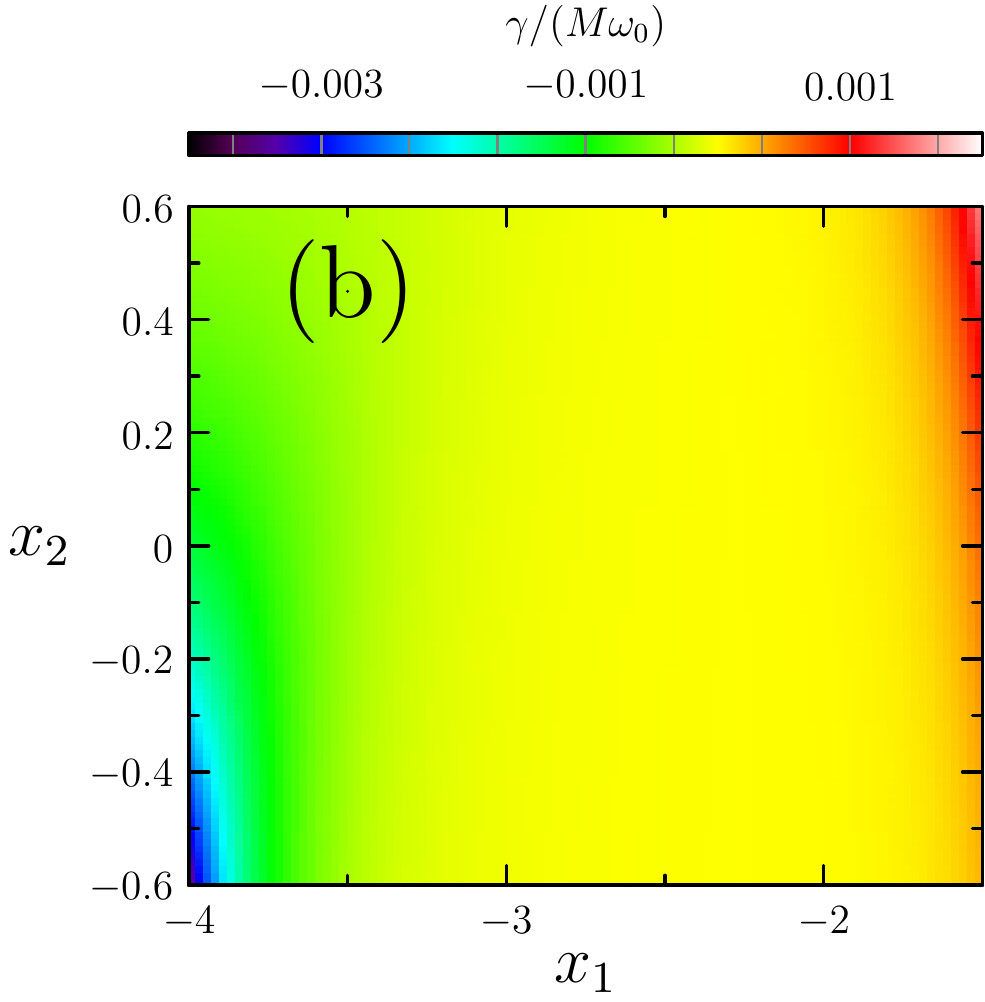}
\end{center}
\caption{Curl of the average force and damping coefficient for the model with two vibrational modes: (a) The curl of the current-induced mean force ${\mathbf F}$ is,
in a non-equilibrium situation, generally non-zero, indicating that the force is non-conservative. (b) One of the two eigenvalues
of $\gamma^s$. Remarkably, it undergoes sign changes. A dissipation matrix $\gamma^s$ which is non-positive definite implies destabilization of the static equilibrium solution found at lower bias potentials, in this case driving the system into a limit cycle, see main text and Fig.
\ref{fig:H2}. The parameters used are such that $\lambda_1/\lambda_2=3/2$. The elastic modes are degenerate with $\hbar \omega_0 = 0.014$, $\Gamma_{L,R}=\frac{1\pm0.8}{2} (\sigma_0 \pm \sigma_z)$, and the hopping between the orbitals is $t=0.9$. The dimensionless coordinates are $x_i = ({M \omega_0^2}/{\lambda}) X_i$ and energies are in units of  ${\lambda^2}/{(M\omega_0^2)}$, where $\lambda=({\lambda_1+\lambda_2})/{2}$.}
\label{fig:H2curl}
\end{figure}

\begin{figure*}
\begin{center} $
\begin{array}{cccc}
\includegraphics[height=4cm,keepaspectratio=true]{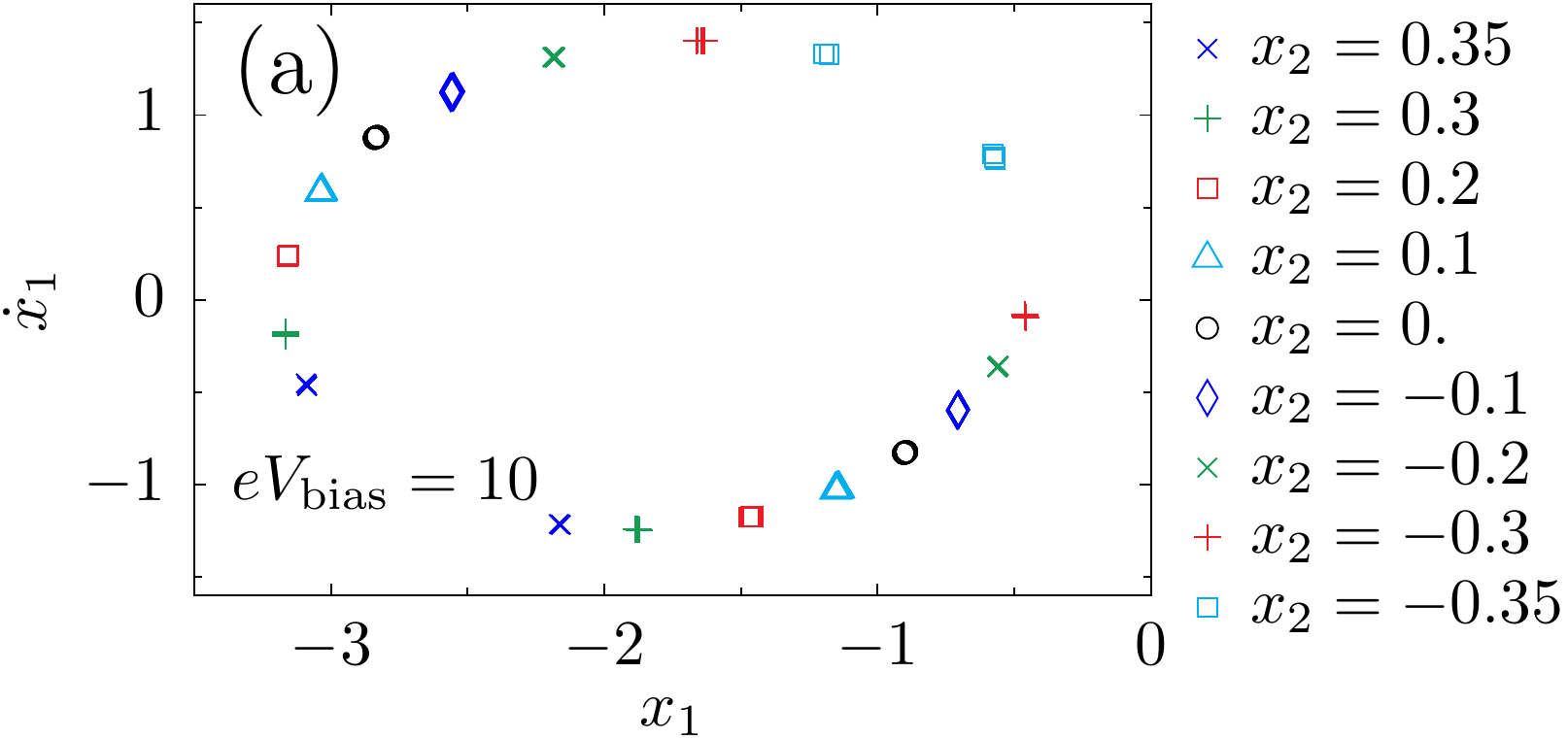} &
\includegraphics[height=4cm,keepaspectratio=true]{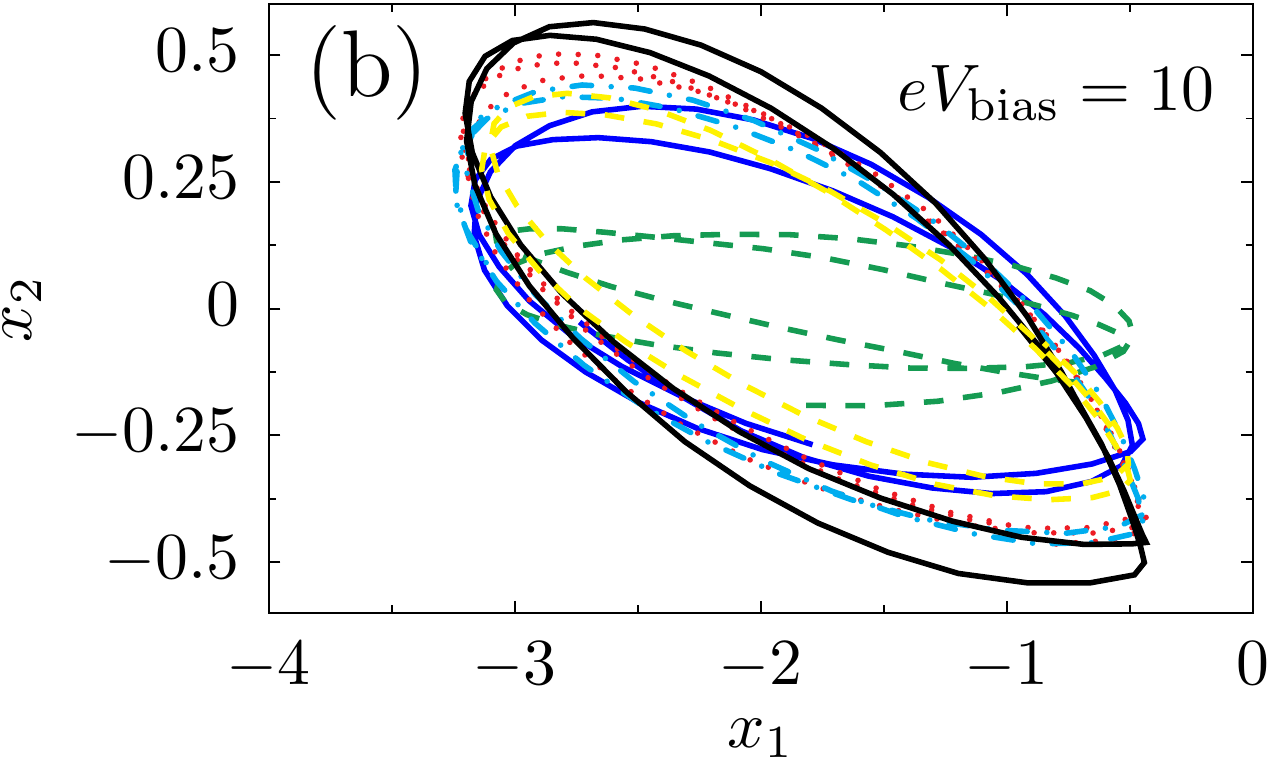}
\end{array}$ 
\end{center}
\caption{Limit-cycle dynamics for the model with two vibrational modes. (a) At large bias voltages, Poincar\'{e} sections of the four dimensional phase space show the presence of a limit cycle in the Langevin dynamics without fluctuating force. (b) Several periods of typical trajectories (for different initial conditions after a transient) in the presence of the fluctuating forces ${\mathbf \xi}$ are shown. The same general parameters as in Fig. \ref{fig:H2curl} are used here.}
\label{fig:H2}
\end{figure*}

We have studied the zero-temperature dynamics of our two-level, two-mode system for different ranges of parameters. In Fig. (\ref{fig:H2curl}) we map out the values of the curl of the mean force, $\left(\nabla\times F\right)_\perp$, indicating that the force is non-conservative throughout parameter space. We also plot one of the two eigenvalues of the dissipation matrix $\gamma^s$, showing that it can take negative values in some regions of the parameter space. We find that it is possible
to drive the system into a limit cycle by varying the bias potential. The existence of this limit cycle is shown in Fig. \ref{fig:H2} (a), where we have plotted various Poincar\'e sections of the non-linear system without fluctuations. The figure shows the trajectory in phase space of the (dimensionless) oscillator coordinate $x_1$ after the dynamical equilibrium is reached, for several cuts of the (dimensionless) coordinate $x_2$. Each cut shows two points in $x_1$ phase  space, indicating the entry and exit of the trajectory. Each point in the plot actually consists of several points that fall on top of each other, corresponding to every time the coordinate $x_2$ has the value indicated in the legend of Fig. \ref{fig:H2} (a). This shows the periodicity of the solution of the non-linear equations of motion for $x_1,\,x_2$ for the particular bias chosen. Surveying over the various values of $x_2$ reveals a closed trajectory in the parametric coordinate space $x_1,\,x_2$. 

Remarkably, signatures of the limit cycle survive the inclusion of the Langevin force. Fig. \ref{fig:H2} (b) depicts typical trajectories in the oscillator's coordinate space $x_1,\,x_2$ in the presence of the stochastic force, showing fluctuating trajectories around the stable limit cycle. 

\begin{figure*}
\begin{center}
\includegraphics[width=8cm,keepaspectratio=true]{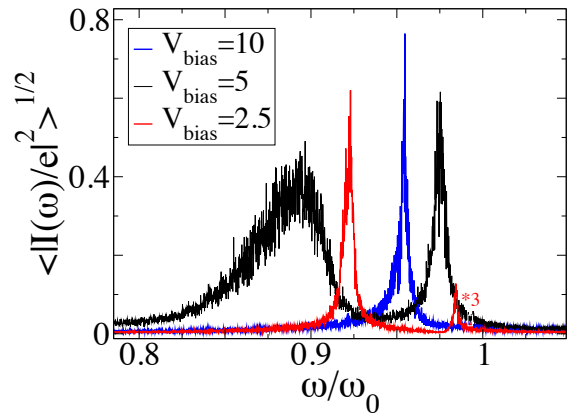}
\end{center}
\caption{Current-current correlation function in the presence of noise for the system with two vibrational modes. The limit cycle is signaled by a single peak ($V_{bias}=10$, see Fig. \ref{fig:H2}), as opposed to two peaks in the absence of a limit cycle ($V_{bias}=2.5,\,5$). Increasing the bias potential increases the noise levels but the peaks are still easily recognizable. The results are obtained by averaging over times long enough compared with the characteristic oscillation times. The same general parameters as in Fig. \ref{fig:H2curl} are used here.}
\label{fig:H2current}
\end{figure*}

Experimentally, the signature of the limit cycle would be most directly reflected in the current-current correlation function, as depicted in Fig. \ref{fig:H2current}. We find that in the absence of a limit cycle the system is dominated by two characteristic frequencies, shown by the peaks in Fig. \ref{fig:H2current}. These frequencies correspond to the shift in energy of the two degenerate vibrational modes due to the average current-induced forces $F_1$ and $F_2$. When the bias voltage is such that the system enters a limit cycle, the current-current correlation shows instead only one peak as a function of frequency. This result, as shown in Fig. \ref{fig:H2current}, is fairly robust to noise, making the onset of limit-cycle dynamics observable in experiment.

\section{Conclusions}
\label{sec5}
Within a non-equilibrium Born-Oppenheimer approximation, the dynamics of a nanoelectromechanical system can be described in terms of a Langevin equation, in which the mechanical modes of the mesoscopic device are subject to current-induced forces. These forces include a mean force, which is independent of velocity and due to the average net force the electrons exert on the oscillator, a stochastic Langevin force which takes into account the thermal and non-equilibrium fluctuations with respect to the mean force value, and a force linear in the velocity of the modes. This last, velocity dependent force, consists of a dissipative term plus a term that can be interpreted as an effective ``Lorentz'' force, due to an effective magnetic field acting in the parameter space of the modes.

In this work we have expressed these current-induced forces through the scattering matrix of the coherent mesoscopic conductor and its parametric derivatives, extending the results found previously in Ref. \bcite{BodePRL11}. Our results are now valid for a generic coupling between the electrons and the vibrational degrees of freedom, given by a matrix $h_0(\mathbf{X})$, and for energy-dependent hybridization with the leads, given by the matrix $W(\epsilon)$. We have shown that expressing {\it all} the current-induced forces in terms of the S-matrix is only possible by going beyond the strictly adiabatic approximation, and it is necessary to include the first order correction in the adiabatic expansion. This introduces a new fundamental quantity into the problem, the A-matrix, which needs to be calculated together with the frozen S-matrix for a given system.

There are several circumstances in which the first non-adiabatic correction, encapsulated in the A-matrix, is necessary. While the average as well as the fluctuating force can be expressed solely in terms of the adiabatic S-matrix, the A-matrix enters both the frictional and the Lorentz-like force. In equilibrium, the frictional force reduces to an expression in terms of the adiabatic S-matrix. Out of equilibrium, however, an important new contribution involving the A-matrix appears. In contrast, the A-matrix is always required  to express the Lorentz-like force, even when the system is in thermal equilibrium. 

The expressions for the current-induced forces in terms of the scattering matrix allow us to extract important properties from general symmetry arguments. Driving the nanoelectromechanical system out of equilibrium by imposing a bias results in qualitatively new features for the forces. We have shown that the mean force is non-conservative in this case, and that the dissipation coefficient acquires a non-equilibrium contribution that can be negative. We have also shown that when considering more than one mechanical degree of freedom, a pseudo Lorentz force is present even for a time-reversal invariant system, unless one also imposes thermal equilibrium on top of the time-reversal condition.

Our model allows one to study, within a controlled approximation, the non-linear dynamics generated by the interplay between current and vibrational degrees of freedom, opening up the path for a systematic study of these devices. By means of simple model examples, we have shown that it is possible to drive a nanoelectromechanical system into interesting dynamically stable regimes such as a limit cycle, by varying the applied bias potential. In a limit cycle, the vibrational modes vary periodically in time, which can be the operating principle for a molecular motor. On the other hand, the possibility of non-conservative forces could also allow one to extract energy from the system, providing a controllable tool for cooling. The study of these kinds of phenomena for realistic systems is an interesting application of the formalism presented in this paper.

\section{Acknowledgments}
We acknowledge discussions with P. W. Brouwer, G. Zarand, and L. Arrachea as well as support by the DFG through SPP 1459, SFB TR/12, and SFB 658.

\newpage
\begin{widetext}
\appendix

\section{Useful relations}
\label{app1}
Here we list a set of useful relations for the derivations in the main text.
\subsection{Green's functions relations}
The Green's functions are related via
\beq\label{GRGAG>G<}
G^R-G^A=G^>-G^<.
\eeq
The lesser and larger Green's functions are given by
\begin{align}
G^< &=G^R\Sigma^<G^A=2 i \sum_\alpha f_\alpha \,G^R \Gamma_\alpha G^A =2\pi i
\sum_\alpha f_\alpha \,G^R W^\dagger \Pi_\alpha WG^A \label{eq:G<ad}\\
G^> &=G^<+G^R-G^A=-2\pi i\sum_\alpha (1-f_\alpha)\, G^R W^\dagger \Pi_\alpha W
G^A. \label{eq:G>ad}
\end{align}
From \eqref{FrozenG} it is easy to see that
\begin{equation}\label{aux}
W^\dagger W = \frac{1}{2\pi i}[ (G^R)^{-1}-(G^A)^{-1} ]\,,
\end{equation}
\beq\label{dgrx}
\partial_{X_\nu} G^{R}= G^{R} \Lambda_\nu G^{R}
\eeq
and
\beq\label{dgre}
\partial_\epsilon G^{R}=-G^{R}(1-\partial_\epsilon\Sigma^R)G^{R}.
\eeq

\subsection{Green's functions and S-matrix relations}
Noting that (for given $t$) 
$\partial_{X_\nu} G^R = G^R \Lambda_\nu G^R$,
we find using Eq.(\ref{aux}):
\begin{equation} \label{sder}
S^\dagger \frac{\partial S}{\partial X_{\nu}} =
 -2\pi i (1+2\pi i W G^A W^\dagger) W G^R \Lambda_\nu
G^R W^\dagger = -2\pi i W G^A \Lambda_\nu G^R  W^\dagger.
\end{equation}
This holds for arbitrary magnitude of $X_\nu$.  

In the main text we use
\begin{align}\label{dxsdaga}
  \frac{1}{\pi}\frac{\partial S^{\dagger}}{\partial X_{\nu}} A_{\nu'} = 2\pi i
W G^A \Lambda_\nu G^A W^\dagger \partial_\epsilon(W G^R) \Lambda_{\nu'} G^R
W^\dagger - W G^A \Lambda_\nu (G^A-G^R) \Lambda_{\nu'} \partial_\epsilon (G^R
W^\dagger),
\end{align}
\begin{align}\label{eq:gsneqderiv}
\pi \left(\left[S^\dagger \frac{\partial S}{\partial X_\nu}, W G^A \Lambda_{\nu'} \frac{\partial(G^R W^\dagger)}{\partial_\epsilon} - \frac{\partial (W G^A)}{\partial_\epsilon} \Lambda_{\nu'} G^R W^\dagger \right]_{-} \right)_s = \left( \frac{\partial S^\dagger}{\partial X_\nu} A_{\nu'} - A^\dagger_{\nu'} \frac{\partial S}{\partial X_\nu} \right)_s
\end{align}
and
\begin{align}\label{sdagdxa}
\left[S^{\dagger} \frac{\partial A_{\nu}}{\partial X_{\nu'}}\right]_a = -2
\pi \left[W
G^A\Lambda_\nu (\partial_\epsilon G^R) \Lambda_{\nu'}G^R W^\dagger \right]_a.
\end{align}

For energy-independent $\Gamma^\alpha$, we can use \eqref{dgre} so that also
\begin{equation}
S^\dagger \frac{\partial S}{\partial \epsilon}  = 2\pi i W G^A G^R W^\dagger,
\end{equation}
\begin{align}\label{desdagdxs}
  \partial_{\epsilon} \left(S^{\dagger} \frac{\partial S}{\partial
X_{\nu}}\right) &= 2\pi i W G^A \left(G^A \Lambda_\nu + \Lambda_\nu G^R\right)
G^R W^{\dagger}
\end{align}
and \eqref{dxsdaga} simplifies to
\begin{align}
  \frac{\partial S^{\dagger}}{\partial X_{\nu}} A_{\nu'} = \pi W G^A
\Lambda_{\nu}
\left(G^A-G^R\right) \left(\Lambda_{\nu'} G^R - G^R \Lambda_{\nu'}\right) G^R W^{\dagger}.
\end{align}

\section{S-matrix derivation of the damping matrix}\label{app2}
The expression for $\gamma^s$ given in Eq. \eqref{sdef} can be written explicitly in terms of retarded and advanced Green's functions as
\beq\label{sdef2}
\gamma^{s}_{\nu\nu'}=2\pi \sum_{\alpha \alpha'}\int d\epsilon f_\alpha {\rm tr} \left\{
\Lambda_\nu G^R W^{\dagger}\Pi_{\alpha} W G^A \Lambda_{\nu'} \partial_\epsilon\left[(1-f_{\alpha'})
 G^R W^{\dagger}\Pi_{\alpha'} W G^A \right]\right \}_s\,.
\eeq
We split Eq. \eqref{sdef2} into two
terms, the first due to the derivative acting
on the Fermi function, the second from the rest, $\gamma^{s}=\gamma^{s (I)}+\gamma^{s (II)}$. The first term is given by
\begin{align}\label{gs1a}
\gamma^{s (I)}_{\nu\nu'} = 2\pi \sum_{\alpha \alpha'}
\int d\epsilon  f_{\alpha} (- \partial_{\epsilon} f_{\alpha'})
{\rm tr} \left\{
\Pi_{\alpha'} W G^A\Lambda_\nu G^R W^{\dagger}\Pi_{\alpha} W G^A \Lambda_{\nu'} 
 G^R W^{\dagger} \right \}_s
\end{align}
where we have used the cyclic invariance of the trace. Similar to the derivation for the mean force, by means of expression \eqref{sder} in Supp. Mat. \ref{app1}, Eq. \eqref{gs1a} can be expressed in terms of the frozen S-matrix as
\begin{align}\label{gsi}
  \gamma^{s (I)}_{\nu\nu'} = - \sum_{\alpha \alpha'}
\int \frac{\rm d\epsilon}{2\pi}  f_{\alpha} (- \partial_{\epsilon}
f_{\alpha'}) {\rm Tr} \left\{ \Pi_\alpha S^\dagger \frac{\partial S}{\partial
X_{\nu}}
\Pi_{\alpha'} S^\dagger \frac{\partial S}{\partial X_{\nu'}} \right\}_s \,.
\end{align}
The second contribution, in terms of $G^R$ and $G^A$, reads
\begin{align}
\gamma^{s (II)}_{\nu\nu'} = (2\pi)^2 \sum_{\alpha \alpha'}
\int \frac{{\rm d}\epsilon}{2\pi}  F_{\alpha\alpha'}
{\rm tr} \left\{
\Lambda_\nu G^R W^\dagger \Pi_\alpha W G^A \Lambda_{\nu'}
\partial_{\epsilon}\left(
 G^R W^\dagger \Pi_{\alpha'} W G^A \right)\right \}_s\,.
\end{align}
It is instructive to split the factor $F_{\alpha \alpha'}$ into
a symmetric and an antisymmetric part under exchange of the lead indices,
$F_{\alpha \alpha'} = F^s_{\alpha \alpha'}+F^a_{\alpha \alpha'}$, with
\beq
\begin{split}
F^s_{\alpha\alpha'}&\equiv \frac12(f_\alpha+f_{\alpha'}-2f_{\alpha}
f_{\alpha'})\\
F^a_{\alpha\alpha'}&\equiv\frac12(f_\alpha-f_{\alpha'})\,.
\end{split}
\eeq
Correspondingly,
we split $\gamma^{s (II)}$ into symmetric $\left[\gamma^{s(IIs)}\right]$ and
antisymmetric $\left[\gamma^{s(IIa)}\right]$ parts in the lead
indices: $\gamma^{s (II)}=\gamma^{s
(IIs)}+\gamma^{s (IIa)}$. Due to its symmetries, $\gamma^{s(IIs)}$ can be easily expressed in terms of the S-matrix, 
\beq\label{gsiis}
\begin{split}
\gamma^{s(IIs)}_{\nu\nu'} &=\pi\sum_{\alpha \alpha'}\int d\epsilon  F^s_{\alpha \alpha'}\partial_\epsilon {\rm tr} \left\{
\Lambda_\nu G^R W^\dagger \Pi_\alpha W G^A \Lambda_{\nu'}
 G^R W^\dagger \Pi_{\alpha'} W G^A \right \}_s\\
&=-\pi\sum_{\alpha \alpha'}\int d\epsilon\left( \partial_\epsilon F^s_{\alpha \alpha'}\right) {\rm tr} \left\{
\Lambda_\nu G^R W^\dagger \Pi_\alpha W G^A \Lambda_{\nu'}
 G^R W^\dagger \Pi_{\alpha'} W G^A \right \}_s\\
&=\frac{1}{4\pi}\sum_{\alpha \alpha'} \int d\epsilon\left( \partial_\epsilon F^s_{\alpha \alpha'}\right) {\rm tr}\left\{\Pi_\alpha S^\dagger \frac{\partial S}{\partial
X_{\nu}}
\Pi_{\alpha'} S^\dagger \frac{\partial S}{\partial X_{\nu'}} \right \}_s
\end{split}
\eeq
where in the second line we have integrated by parts since $F^{s}$ vanishes for
$\epsilon\to \pm \infty$, and in the last line we have used Eq. \eqref{sder} from App. \ref{app1} once again.

\subsection{``Equilibrium'' dissipative term $\gamma^{s,eq}$}
Since in equilibrium $F^a_{\alpha\alpha'}=
F^a_{\alpha\alpha}=0$, $\left.\gamma^{s(IIa)}\right|_{eq}=0$ and we can
now regroup terms into an ``equilibrium'' contribution,
$\gamma^{s,eq}=\gamma^{s (I)}+\gamma^{s
(IIs)}$, and a purely non-equilibrium contribution
$\gamma^{s,ne}\equiv\gamma^{s (IIa)}$:
\beq
\gamma^{s}=\gamma^{s,eq}+\gamma^{s,ne}\,.
\eeq
By adding up expressions \eqref{gsi} and \eqref{gsiis}, it is straightforward to obtain Eq. \eqref{gseq} for $\gamma^{s,eq}$ given in the main text. 

\subsection{Non-equilibrium dissipative term $\gamma^{s,ne}$}
To obtain $\gamma^{s,ne}$ in terms of S-matrix quantities we start from the expression
\begin{align}\label{gsne0}
\gamma^{s,ne}_{\nu\nu'}= 2\pi \sum_{\alpha \alpha'}
\int  {\rm d}\epsilon F_{\alpha\alpha'}^a 
{\rm tr} \left\{
\Lambda_\nu G^R W^\dagger \Pi_\alpha W G^A \Lambda_{\nu'}
\partial_{\epsilon}\left(
 G^R W^\dagger \Pi_{\alpha'} W G^A \right)\right \}_s\,,
\end{align}
and exploiting $\sum_\alpha \Pi_\alpha=1$ and the identity (\ref{sder}) in Supp. Mat. \ref{app1}, we note that Eq. \eqref{gsne0} can be written as
\begin{align}\label{gsne1}
\gamma^{s,ne}_{\nu\nu'} = -\frac{i}{2}\int  {\rm d}\epsilon\sum_\alpha f_\alpha{\rm tr} \left\{\Pi_\alpha\left[S^\dagger \frac{\partial S}{\partial X_\nu}, W G^A \Lambda_{\nu'} \frac{\partial(G^R W^\dagger)}{\partial\epsilon} - \frac{\partial (W G^A)}{\partial\epsilon} \Lambda_{\nu'} G^R W^\dagger \right] \right\}_s \,,
\end{align}
where $[.\,,\,.]$ indicates the commutator. Calculating each term in the commutator separately we obtain
\begin{widetext}
\beq
\begin{split}
S^\dagger \frac{\partial S}{\partial X_\nu}\left[W G^A \Lambda_{\nu'} \frac{\partial(G^R W^\dagger)}{\partial\epsilon} - \frac{\partial (W G^A)}{\partial\epsilon} \Lambda_{\nu'} G^R W^\dagger\right]&=-WG^A\Lambda_\nu(G^A-G^R)\Lambda_{\nu'}\frac{\partial(G^RW^\dagger)}{\partial\epsilon}\\
&+2\pi i WG^A\Lambda_\nu G^R W^\dagger\frac{\partial(WG^A)}{\partial\epsilon}\Lambda_{\nu'}G^RW^\dagger\\
\left[W G^A \Lambda_{\nu'} \frac{\partial(G^R W^\dagger)}{\partial\epsilon} - \frac{\partial (W G^A)}{\partial\epsilon} \Lambda_{\nu'} G^R W^\dagger\right]S^\dagger \frac{\partial S}{\partial X_\nu}&=-\frac{\partial(WG^A)}{\partial\epsilon}\Lambda_{\nu'}(G^A-G^R)\Lambda_{\nu}G^RW^\dagger\\
&-2\pi i WG^A\Lambda_{\nu'}\frac{\partial( G^R W^\dagger)}{\partial\epsilon}WG^A\Lambda_{\nu}G^RW^\dagger\,,
\end{split}
\eeq
\end{widetext}
where we have used Eq. \eqref{aux} from Supp. Mat. \ref{app1}. Finally, with help of the identity \eqref{dxsdaga} in Supp. Mat. \ref{app1}, the
non-equilibrium term can be expressed as Eq. \eqref{gammaneqS} in the main text.

\section{Resonant level forces: alternative expressions}\label{app23}
To calculate the current-induced forces for the resonant level model presented in Sec. \ref{sec4}, we can alternatively start with the popular S-matrix parametrization ~\bcite{NazarovBook,BennettPRL10}
\begin{equation}\label{singlechannel}
S =  \left( \begin{array}{cc} \sqrt{1-{\cal T}} e^{i\theta} & \sqrt{{\cal T}}
e^{i\eta} \\ \sqrt{{\cal T}} e^{i\eta} & 
-\sqrt{1-{\cal T}} e^{i(2\eta-\theta)}
\end{array}\right)\,,
\end{equation}
where the transmission coefficient ${\cal T}$ and the phases $\eta,\theta$ depend on $X$. We present here the results for linear coupling, $\tilde\epsilon(X)=\epsilon_0+\lambda X$. We can then identify the transmission probability
\begin{equation}
{\cal T}(\epsilon,X) = \frac{4\Gamma_L\Gamma_R}{(\epsilon-\epsilon_0-\lambda
X)^2+\Gamma^2} 
\end{equation}
and the phases
\begin{eqnarray*}
\eta(\epsilon,X) &=& -\frac{\pi}{2} - \arctan
 \left(\frac{\Gamma}{\epsilon-\epsilon_0-\lambda X}\right)
\\
\theta(\epsilon,X) &=& \frac{\pi}{2} + \eta + \arctan \left( \frac{\Gamma_R-\Gamma_L}
{\epsilon-\epsilon_0-\lambda X} \right)\,.
\end{eqnarray*}
We can now relate the current-induced forces to this S-matrix parametrization. The result for the average force can be split into a non-equilibrium force
$F^{ne}$ and an equilibrium force $F^{eq}$, {\it i.e.}, $F=F^{ne}+F^{eq}$
with
\begin{eqnarray}
F^{ne}(X) & = &  \int \frac{d\epsilon}{2\pi} (f_L-f_R)  (1-{\cal T})
\frac{\partial(\theta-\eta)}{\partial X}  \\ 
\nonumber 
F^{eq}(X)&=& \int \frac{d\epsilon}{2\pi} (f_L+f_R) \frac{\partial\eta}
{\partial X}\,.
\end{eqnarray}
The amplitude of the fluctuating force can be obtained from Eq. \eqref{variance2} and is given by
\begin{equation}
D(X)= \int\frac{d\epsilon}{2\pi} \sum_{\alpha\alpha'}
 F^s_{\alpha\alpha'} Y_{\alpha\alpha'} \,,
\end{equation} 
where we have defined
\begin{eqnarray*}
Y_{LL} &=&  \left[(1-{\cal T}) 
\frac{\partial(\eta-\theta)}{\partial X}-
\frac{\partial\eta}{\partial X} \right ]^2   \\ 
Y_{RR} &=&  \left[(1-{\cal T}) 
\frac{\partial(\eta-\theta)}{\partial X}+ 
\frac{\partial\eta}{\partial X} \right ]^2   \\ 
Y_{LR} &=& Y_{RL} = \frac{1}{4{\cal T}(1-{\cal T})} 
\left(\frac{\partial {\cal T}}{\partial X}\right)^2 
+ {\cal T}(1-{\cal T}) \left(\frac{\partial(\eta-\theta)}{\partial X}\right)^2.
\end{eqnarray*}
After some algebra, we also obtain
\begin{equation}
\gamma^{s}(X) =\frac{1}{2T}\left[ D(X) -  \int\frac{d\epsilon}{2\pi} (f_L-f_R)^2 Y_{LR}\right]\,.
\end{equation}
This last expression corresponds to $\gamma^{s,eq}$ given in Eq. \eqref{gseq}. (As we pointed out previously, $\gamma^{s,ne}$ vanishes in this case). Here we have isolated a term that vanishes in equilibrium, showing explicitly that there is a non-equilibrium contribution in \eqref{gseq}.

\section{Current-induced forces for the two-level model}\label{app23b}
The mean force is given by
\begin{widetext}
\beq\label{MeanF2L}
F(X)= -\lambda_1 \Gamma \int \frac{d \epsilon}{2\pi} \left[(f_L+f_R) \frac{2
\lambda_1 X (\epsilon-\epsilon_0)}{\left\vert \Delta\right\vert^2} + (f_L-f_R)
\frac{(\epsilon-\epsilon_0)^2 +(\lambda_1 X)^2 -t^2 + (\Gamma/2)^2}{\left\vert \Delta\right\vert^2}\right]\,.
\eeq
\end{widetext}
The friction coefficient $\gamma^s=\gamma^{s,eq}+\gamma^{s,ne}$ reads
\begin{widetext}
\begin{align}\label{Gamma2L}
  \gamma^{s,eq} =& \frac{\lambda_1^2 \Gamma^2}{4 \pi}
\int d\epsilon \left\{-\frac{\partial_\epsilon f_L+\partial_\epsilon
f_R}{\left\vert \Delta\right\vert^4}
\Bigl[\left((\epsilon-\epsilon_0)^2 + (\Gamma/2)^2+(\lambda_1 X)^2 +t^2\right)^2 +
\left(2(\epsilon-\epsilon_0) \lambda_1 X \right)^2 \right.\nonumber\\
  -&\left. \left(2
(\epsilon-\epsilon_0) t\right)^2\Bigr] +\frac{\partial_\epsilon
f_R-\partial_\epsilon f_L}{\left\vert \Delta\right\vert^4} \left[4 (\epsilon-\epsilon_0) \lambda_1 X
\left((\epsilon-\epsilon_0)^2 +(\Gamma/2)^2 +(\lambda_1 X)^2 - t^2\right)
\right]\right\}\nonumber\,,\\
  \gamma^{s,ne} =& \frac{2 \lambda_1^2 \Gamma^2 t^2 \lambda_1 X}{\pi}
\int d\epsilon\,
\frac{f_R - f_L}{\left\vert \Delta\right\vert^6} \Bigr[\left((\epsilon-\epsilon_0)^2-(\lambda_1 X)^2 - t^2\right)^2 \nonumber\\+& 2 (\Gamma/2)^2 \left((\epsilon-\epsilon_0)^2+(\lambda_1 X)^2 + t^2\right) +(\Gamma/2)^4 \Bigl]\,.
\end{align}
\end{widetext}
\section{Current-induced forces for the two vibrational modes model}\label{app3}
Here we list the current-induced forces quantities, calculated from Eqs. \eqref{force}, \eqref{variance2}, \eqref{eq:damp2} and \eqref{effB2} for the two-modes example discussed in the main text. For convenience, we define the following quantities:
\begin{eqnarray}
g_{\alpha0}(\epsilon) & = & \frac{(\epsilon-\tilde{\epsilon})^{2}+\tilde{t}^{2}+\Gamma_{1-\alpha}^{2}}{\left|\Delta\right|^{2}}\\
g_{\alpha1}(\epsilon) & = & \frac{2\tilde{t}\,(\epsilon-\tilde{\epsilon})}{\left|\Delta\right|^{2}}\\
g_{\alpha2}(\epsilon) & = & \pm\frac{-2\tilde{t}\,\Gamma_{1-\alpha}}{\left|\Delta\right|^{2}}\\
g_{\alpha3}(\epsilon) & = & \pm\frac{(\epsilon-\tilde{\epsilon})^{2}+\Gamma_{1-\alpha}^{2}-\tilde{t}^{2}}{\left|\Delta\right|^{2}}
\end{eqnarray}
where the $+(-)$ refers to $\alpha=L(R)$ and with $1-\alpha=R(L)$
for $\alpha=L(R)$, and  $\Delta(X_1,X_2)=(\epsilon-\tilde\epsilon+i\Gamma_L)(\epsilon-\tilde\epsilon+i\Gamma_R)-\tilde t^2$.
\subsection{Mean force}
\begin{eqnarray}
F_{1} & = & -2\int\frac{d\epsilon}{2\pi}\,\lambda_{1}\sum_{\alpha}\frac{f_{\alpha}(\epsilon)\Gamma_{\alpha}\left((\epsilon-\tilde{\epsilon})^{2}+\tilde{t}^{2}+\Gamma_{1-\alpha}^{2}\right)}{\left[(\epsilon-\tilde{\epsilon})^{2}-\tilde{t}^{2}-\Gamma_{L}\Gamma_{R}\right]^{2}+\left[(\Gamma_{L}+\Gamma_{R})(\epsilon-\tilde{\epsilon})\right]^{2}}\\
F_{2} & = & -4\int\frac{d\epsilon}{2\pi}\,\lambda_{2}\frac{\tilde{t}\,(\epsilon-\tilde{\epsilon})\,\left(f_{L}(\epsilon)\Gamma_{L}+f_{R}(\epsilon)\Gamma_{R}\right)}{\left[(\epsilon-\tilde{\epsilon})^{2}-\tilde{t}^{2}-\Gamma_{L}\Gamma_{R}\right]^{2}+\left[(\Gamma_{L}+\Gamma_{R})(\epsilon-\tilde{\epsilon})\right]^{2}}
\end{eqnarray}
\subsection{Fluctuating force}
\begin{eqnarray}
D_{11} & = & 2\left(\lambda_{1}\right)^{2}\int\frac{d\epsilon}{2\pi}\sum_{\alpha\beta}f_{\alpha}(\epsilon)\Gamma_{\alpha}\left(1-f_{\beta}(\epsilon)\right)\Gamma_{\beta}\sum_{\mu}g_{\alpha\mu}g_{\beta\mu}\\
D_{12} & = & 2\lambda_{1}\lambda_{2}\int\frac{d\epsilon}{2\pi}\sum_{\alpha\beta}f_{\alpha}(\epsilon)\Gamma_{\alpha}\left(1-f_{\beta}(\epsilon)\right)\Gamma_{\beta}\left(g_{\alpha0}g_{\beta1}+g_{\alpha1}g_{\beta0}\right)\\
D_{22} & = & 2\left(\lambda_{2}\right)^{2}\int\frac{d\epsilon}{2\pi}\sum_{\alpha\beta}f_{\alpha}(\epsilon)\Gamma_{\alpha}\left(1-f_{\beta}(\epsilon)\right)\Gamma_{\beta}\left(g_{\alpha0}g_{\beta0}+g_{\alpha1}g_{\beta1}-g_{\alpha2}g_{\beta2}-g_{\alpha3}g_{\beta3}\right)
\end{eqnarray}
\subsection{Damping coefficients}
\begin{eqnarray}
\gamma_{11}^{s}& = & \frac{\left(\lambda_{1}\right)^{2}}{2\pi}\int d\epsilon\sum_{\alpha\beta}\left(-\partial_{\epsilon}f_{\alpha}(\epsilon)\right)\Gamma_{\alpha}\Gamma_{\beta}\,\sum_{\mu}g_{\alpha\mu}g_{\beta\mu}\\
\gamma_{12}^{s} & = & 2\lambda_{1}\lambda_{2}\int\frac{d\epsilon}{2\pi}\sum_{\alpha\beta} f_{\alpha}(\epsilon)\Gamma_{\alpha}\left(-\partial_{\epsilon}f_{\beta}(\epsilon)\right)\Gamma_{\beta}\left(g_{\alpha0}g_{\beta1}+g_{\alpha1}g_{\beta0}\right)\\
\gamma_{22}^{s} & = & 2\left(\lambda_{2}\right)^{2}\int\frac{d\epsilon}{2\pi}\sum_{\alpha\beta} f_{\alpha}(\epsilon)\Gamma_{\alpha}\left(-\partial_{\epsilon}f_{\beta}(\epsilon)\right)\Gamma_{\beta}\left(g_{\alpha0}g_{\beta0}+g_{\alpha1}g_{\beta1}-g_{\alpha2}g_{\beta2}-g_{\alpha3}g_{\beta3}\right)
\end{eqnarray}
\subsection{``Lorentz'' term}    
\beq
\gamma_{12}^{a}=  -2\tilde{t}\,\frac{\lambda_{1}\lambda_{2}}{\pi}\Gamma_{L}\Gamma_{R}(\Gamma_{L}^{2}-\Gamma_{R}^{2})\int d\epsilon\left[\partial_{\epsilon}\frac{\epsilon-\tilde{\epsilon}}{\left|\Delta\right|^{2}}\right]\left[\frac{f_{L}-f_{R}}{\left|\Delta\right|^{2}}\right]
\eeq
\end{widetext}

\bibliography{LongCIF}

\end{document}